\documentclass[twocolumn,english,aps,prb,floats]{revtex4}
\usepackage[T1]{fontenc}
\usepackage[latin9]{inputenc}
\usepackage{babel}
\usepackage{textcomp}
\usepackage{amsmath}
\usepackage{amssymb}
\usepackage{graphicx}
\usepackage{esint}
\usepackage[unicode=true,pdfusetitle,
 bookmarks=true,bookmarksnumbered=false,bookmarksopen=false,
 breaklinks=false,pdfborder={0 0 1},backref=false,colorlinks=false]
 {hyperref}

\makeatletter

\newcommand{\lyxmathsym}[1]{\ifmmode\begingroup\def\b@ld{bold}
  \text{\ifx\math@version\b@ld\bfseries\fi#1}\endgroup\else#1\fi}

\@ifundefined{textcolor}{}
{%
 \definecolor{BLACK}{gray}{0}
 \definecolor{WHITE}{gray}{1}
 \definecolor{RED}{rgb}{1,0,0}
 \definecolor{GREEN}{rgb}{0,1,0}
 \definecolor{BLUE}{rgb}{0,0,1}
 \definecolor{CYAN}{cmyk}{1,0,0,0}
 \definecolor{MAGENTA}{cmyk}{0,1,0,0}
 \definecolor{YELLOW}{cmyk}{0,0,1,0}
}

\usepackage{ifpdf}
\usepackage{babel}
\usepackage{bm}
\usepackage{wasysym}
\usepackage{breakurl}

\@ifundefined{textcolor}{}{%
 \definecolor{BLACK}{gray}{0}
 \definecolor{WHITE}{gray}{1}
 \definecolor{RED}{rgb}{1,0,0}
 \definecolor{GREEN}{rgb}{0,1,0}
 \definecolor{BLUE}{rgb}{0,0,1}
 \definecolor{CYAN}{cmyk}{1,0,0,0}
 \definecolor{MAGENTA}{cmyk}{0,1,0,0}
 \definecolor{YELLOW}{cmyk}{0,0,1,0}
}

\makeatother

\begin{document}
\title{Random Anisotropy Magnet at Finite Temperature}
\author{Dmitry A. Garanin and Eugene M. Chudnovsky}
\affiliation{Physics Department, Herbert H. Lehman College and Graduate School,
The City University of New York, 250 Bedford Park Boulevard West,
Bronx, New York 10468-1589, USA }
\date{\today}
\begin{abstract}
We present finite-temperature Monte Carlo studies of a 2D random-anisotropy
magnet on lattices containing one million spins. The correlated spin-glass
state predicted by analytical theories is reproduced in simulations,
as are the field-cooled and zero-field-cooled magnetization curves
observed in experiments. The orientations of lattice spins begin to
freeze when temperature is lowered. The freezing transition is due
to the energy barriers generated by the random anisotropy rather than
due to random interactions in conventional spin-glasses. We describe
freezing by introducing the time-dependent spin-glass order parameter
$q$ and the spin-melting time $\tau_{M}$ defined via $q=\tau_{M}/t$
above freezing, where $t$ is the time of the experiment represented
by the number of Monte Carlo steps.
\end{abstract}
\maketitle

\section{Introduction}

Amorphous and nanocrystalline ferromagnets have multiple technological
applications due to their remarkable magnetic softness \citep{Marin-MMM2020}.
Many of such systems are characterized by ferromagnetic exchange and
random local magnetic anisotropy. They received the name of random-anisotropy
(RA) ferromagnets. Their static properties have been intensively studied
in the past, see, e.g., Refs.\ \onlinecite{RA-book,CT-book,PCG-2015}
and references therein. Recently, it has been shown that RA magnets
can also be excellent broadband absorbers of microwave radiation \citep{GC-PRB2021}.

Theoretical research on RA magnets received strong initial boost from
a seminal work of Imry and Ma \citep{IM} who argued that random on-site
field of strength $h$, no matter how weak, destroys ferromagnetic
order exponentially fast beyond the distance $R_{f}$ that scales
as $(J/h)^{2/(4-d)}$, where $J$ is the exchange constant and $d=1,2,3$
is the dimensionality of the system. This gave rise to the concept
of Imry-Ma (IM) domains of average size $R_{f}$, representing a system
in which local direction of magnetization wanders smoothly on a scale
$R_{f}$, resulting in a zero net magnetization of a large system.
Although random anisotropy of strength $D_{R}$ is different from
the random field, it does generate random effective field at the lattice
site, which makes plausible the picture of IM domains of size $R_{f}\sim(J/D_{R})^{2/(4-d)}$
(in lattice units) in that case, too. This magnetic state received
the name of the correlated spin glass (CSG) \citep{EC-1983,CSS-1986}.

The CSG theory explained many features of amorphous magnets observed
in experiments. Conceptually similar models were developed for arrays
of magnetic bubbles \citep{bubbles}, vortex lattices in superconductors
\citep{EC-PRB1989,Blatter-RMP1994}, charge-density waves \citep{Efetov-77,Okamoto-2015,PC-PRB2015},
liquid crystals \citep{LC}, and He-3 in aerogel \citep{Volovik-JLTP2008,Li-Nature2013}
and on corrugated graphene \citep{Volovik-JETPlett2018}. Later on,
the validity of the concept of IM domains was questioned by people
who applied the renormalization group theory and replica symmetry
breaking methods to the RA model and to the equivalent model of pinned
flux lattices in superconductors, see, e.g., Refs. \onlinecite{Feldman-2000,Nattermann-2000}
and references therein. The Bragg-glass phase characterized by the
power-law decay of correlations instead of exponential decay was proposed,
but that prediction was never confirmed by any experiment on magnetic
systems.

Another criticism of the IM concept came from its neglect of metastable
states \citep{SL-JAP1987,DB-PRB1990,DC-1991}. It was found numerically
that the RA magnets exhibited metastability and history dependence
\citep{nonergodic,GC-EPJ}, although they do break into IM domains
of size predicted by theory if one begins with a fully disordered
initial state. It was demonstrated that the relation between the number
of spin components and dimensionality of space in the random-field
model determines whether the model possesses topological defects,
and that the latter is crucial for preservation or decay of the long-range
correlations \citep{GCP-PRB2013,PGC-PRL,CG-PRL}.

The RA model turned out to be more challenging than the random-field
model. Its exact ground state, spin-spin correlation functions, and
classification of topological defects has never been established with
certainty despite the significance of RA magnets for applications.
Previous analytical work on small lattices, accompanied by Monte Carlo
studies, that assumed thermal equilibrium \citep{Fisch,Itakura,Imagawa,Dudka},
could not describe time-dependent behavior and hysteresis observed
in real systems. The hysteresis curve and scaling of coercivity arising
from the presence of topological defects in a 3D RA model have been
studied numerically in Ref. \ \onlinecite{PCG-2015}. Scaling arguments
were developed that helped understand numerical results.

In this article, with the help of the Monte Carlo technique, we address
temporal behavior of RA systems as it is usually done for spin glasses.
In particular, we study melting of spin states, that has been seldom
investigated theoretically so far. We study the evolution (in terms
of Monte Carlo steps) of the RA magnet at different temperatures,
starting with the quenched state with a random orientation of spins.
This kind of numerical experiment mimics preparation of an amorphous
magnet from a disordered paramagnetic state by a melt spinning technique
\citep{Marin-MMM2020}. It helps to answer a long-standing question
\citep{Binder} whether on lowering temperature the RA magnet undergoes
freezing of correlated spin groups due to energy barriers created
by the local magnetic anisotropy or it exhibits a spin-glass transition
due to interaction between correlated spin groups.

Theoretical \citep{Billoni-PRB2005} and experimental \citep{Shand-JAP2005}
studies of that problem so far have addressed systems with large RA
compared to the exchange, when individual spins behave similar to
single-domain magnetic particles and extended ferromagnetic correlations
are absent. Here we study the less obvious limit of a soft magnet
in which the RA of the order of the exchange or smaller. It is the
case of the CSG with extended ferromagnetic correlations and large
magnetic susceptibility.

To study spin correlations in the CSG one needs a large system. The
power of modern computers is better suited for that task than it was
in the past. Still the 3D case requires impractically large computational
times, so we stick to a 2D system of one million spins. The paper
is organized as follows. The model and properties of the CSG that
follow from the IM argument are discussed in Section II. The freezing
parameter and melting time that describe physical properties of the
system, together with formulas for magnetization and susceptibility,
are introduced in Section III. Our numerical method is described in
Section IV. Numerical results on field-cooled and zero-field-cooled
magnetization curves are presented in Section V-A. The computed temperature
dependence of the freezing parameter and the melting time is given
in Section V-B. Results on the magnetization and susceptibility are
included in Section V-C. The final Section VI contains discussion
of the nature of the observed freezing transition.

\section{The model }

\label{sec:The-model}

We consider the model of a classical random-anisotropy (RA) ferromagnet
on a lattice

\begin{equation}
\mathcal{H}=-\frac{1}{2}\sum_{ij}J_{ij}\mathbf{s}_{i}\cdot{\bf s}_{j}-\frac{D_{R}}{2}\sum_{i}({\bf n}_{i}\cdot{\bf s}_{i})^{2}-\mathbf{H}\cdot\sum_{i}{\bf s}_{i}.\label{Hamiltonian}
\end{equation}
Here $J_{ij}$ is the nearest-neighbor coupling of the classical spin
vectors $\left|{\bf s}_{i}\right|=1$ with the coupling constant $J>0$,
$D_{R}$ is the RA constant, ${\bf n}_{i}$ are randomly oriented
easy-axis vectors, and $\mathbf{H}$ is the external field in the
energy units. This model shares many features with spin glasses. At
low temperatures for $H=0$, spins tend to locally order in the directions
of the locally predominant orientation of the anisotropy axis. For
$D_{R}/J\lesssim1$ there is a strong short-range order as the ferromagnetic
correlation radius\citep{RA-book}
\begin{equation}
R_{f}\sim a\left(\frac{J}{D_{R}}\right)^{2/\left(4-d\right)}\label{Rf}
\end{equation}
 becomes much larger than the lattice spacing $a$. There is a numerical
factor of about 10 in this formula, see the estimations below Eq.
(\ref{Rf_via_m}). At low temperatures, the magnetic structure consists
of large correlated regions in which spins point in the direction
of the predominant anisotropy that is random. Such correlated regions
can be called ``Imry-Ma domains'' (IM domains), although there are
no domain walls between them. The correlation radius is especially
large in three dimensions, $d=3$. The result for $R_{f}$ above can
be obtained with the help of the Imry-Ma argument. Suppose the spins
are correlated within the distance $R_{f}$. Averaging the RA energy
over this region gives the energy
\begin{equation}
E_{RA}\sim-D_{R}\left(\frac{a}{R_{f}}\right)^{d/2}\label{ERA}
\end{equation}
per spin for $a\ll R_{f}$. The exchange energy per spin due to the
change of the spin field at the distance $R_{f}$ is
\begin{equation}
E_{ex}\sim J\left(\frac{a}{R_{f}}\right)^{2}.\label{Eex}
\end{equation}
Minimizing the total energy $E_{tot}=E_{RA}+E_{ex}$ with respect
to $R_{f}$ yields Eq. (\ref{Rf}). For this $R_{f}$, both anisotropy
and exchange energies have the same order of magnitude, $\left|E_{RA}\right|\sim E_{ex}$.
This picture assumes that the spins within IM domains are directed
along the anisotropy axis averaged over the IM domain.

One can estimate the zero-field zero-temperature susceptibility of
the RA magnet as follows. \citep{CS-1984,CSS-1986} If a small uniform
field $H$ is applied, the spins deviate from the dominant-anisotropy
direction by a small angle $\delta\theta$, that results in the energy
change $\delta E\sim-H\delta\theta+\left|E_{RA}\right|\left(\delta\theta\right)^{2}$.
Minimizing this energy with respect to $\delta\theta$ and using $\left|E_{RA}\right|\sim E_{ex}$,
for the susceptibility in the energy units $\chi\sim\delta\theta/H$
one obtains
\begin{equation}
\chi=\frac{k}{J}\left(\frac{R_{f}}{a}\right)^{2}.\label{chi_estimation}
\end{equation}
where $k$ is a factor of order unity. The latter depends on the exact
form of the spin-spin correlation function. In 2D this factor also
contains logarithmic dependence on $R_{f}$. At $R_{f}\gg a$ the
susceptibility is large, which explains magnetic softness of RA magnets.

As it was mentioned above, the correlated bunches of spins (IM domains)
tend to orient themselves in the two possible directions along the
predominant anisotropy axis. The energy barrier $\Delta U$ between
these orientations can be estimated as $\Delta U\sim E_{RA}$, Eq.
(\ref{ERA}). Using Eq. (\ref{Rf}), one obtains
\begin{equation}
\Delta U\sim D_{R}\left(\frac{J}{D_{R}}\right)^{d/\left(4-d\right)}=J\left(\frac{D_{R}}{J}\right)^{\frac{2(2-d)}{4-d}}.\label{DeltaU}
\end{equation}
In particular,
\begin{equation}
\Delta U\sim J\begin{cases}
\left(\frac{D_{R}}{J}\right)^{2/3}, & d=1\\
1, & d=2\\
\left(\frac{J}{D_{R}}\right)^{2}, & d=3.
\end{cases}\label{DeltaU_cases}
\end{equation}

Bunches of spins of the size $R_{f}$ that flip over the barrier are
not independent but interacting with their neighbors via the exchange.
The interaction energy can be estimated assuming that the distance
between the neighboring regions of correlated spins is $R_{f}$, so
that the overlap volume is $R_{f}^{d}$. The interaction energy then
has the same form as the exchange energy in the IM argument:
\begin{equation}
E_{int}\sim J\left(\frac{a}{R_{f}}\right)^{2}\left(\frac{R_{f}}{a}\right)^{d}=J\left(\frac{R_{f}}{a}\right)^{d-2}\sim\Delta U.\label{E_int}
\end{equation}
That is, flipping bunches of correlated spins are strongly coupled.
Thus the RA magnet has a similarity with an ensemble of interacting
magnetic particles with a random anisotropy. However, this analogy
is incomplete as IM domains are not real domains and the boundaries
between these ``particles'' are washed out. The fact that the interaction
between IM domains is comparable with their effective anisotropy energy
makes the situation more complicated. Some of IM domains can be directed
along their effective anisotropy axes while some cannot because of
the interaction with their neighbors. There should be many different
ways to minimize the energy with different sets of ``lucky'' and
``unlucky'' IM domains.

The time required to overcome the collective energy barrier for a
large number of correlated spins should be very long, so that in the
intermediate temperature range the system does not come to equilibrium
during sustainable simulation times. At higher temperatures, transitions
between different states are faster and the system reaches the full
(global) equilibrium. At lower temperatures, spin bunches cannot overcome
energy barriers at all. Here, the local equilibrium near one of the
many local energy minima of the system is established relatively fast.

In finite-size systems with linear size $L$, the results above are
valid for $R_{f}\lesssim L$. The value of the random anisotropy at
which $R_{f}\sim L$ can be estimated as
\begin{equation}
D_{R}\sim D_{R}^{*}=J\left(\frac{a}{L}\right)^{(4-d)/2}.
\end{equation}
For $D_{R}\lesssim D_{R}^{*}$ the barrier can be estimated as
\begin{equation}
\Delta U\sim D_{R}\left(\frac{L}{a}\right)^{d/2}
\end{equation}
that must be smaller than the value for the infinite system. Upon
increasing $L$, the barrier approaches its limiting value from below.

\section{The spin-glass order parameter and other computable quantities}

\label{sec:The-spin-glass-order}

The indicator of the glassy transition is freezing described by the
time autocorrelation function averaged over all $N$ spins:
\begin{equation}
K(\tau)=\frac{1}{N}\sum_{i=1}^{N}\mathbf{s}_{i}(t)\cdot\mathbf{s}_{i}(t+\tau).\label{K_def}
\end{equation}
If the system is at global or local equilibrium, the result does not
depend on the time $t$. However, in the intermediate temperature
interval the system evolves in the direction of equilibrium but cannot
reach it during the observation (simulation) time, thus the result
also depends on $t$. In the glassy state, the spins are frozen and
do not deviate much from their initial positions, so that $K(\tau)$
is finite at large $\tau$. Above the glassy transition, spins are
fluctuating wildly, so that $K(\tau)\rightarrow0$ at large $\tau$.
For large systems, computation of $K(\tau)$ is prohibitive as it
requires keeping all spin configurations in memory over a long time
interval.

The SG order parameter based on the temporal evolution of spins can
be defined as
\begin{equation}
q=\frac{1}{N}\sum_{i=1}^{N}\left\langle \mathbf{s}_{i}\right\rangle _{t}\cdot\left\langle \mathbf{s}_{i}\right\rangle _{t},\label{q_def}
\end{equation}
where
\begin{equation}
\left\langle \mathbf{s}_{i}\right\rangle _{t}\equiv\frac{1}{t_{\max}}\intop_{0}^{t_{\max}}dt\mathbf{s}_{i}(t)\label{s_avr_t}
\end{equation}
is the time average over a long time interval. In spin glasses below
the freezing point, $q\rightarrow\mathrm{const}$ for $t_{\max}\rightarrow\infty$.
This definition is similar to Eq. (1.4) of Ref. \citep{Binder}. Instead
of the time averaging or ensemble averaging we use averaging over
statistical samples generated by the Monte Carlo process. Averaging
over realizations of the RA was done for smaller systems but it was
found that it is better to consider larger systems without this averaging
as large systems self-average.

Above the freezing point, the glassy CF $K(\tau)$ asymptotically
vanishes and one can rewrite $q$ as
\begin{equation}
q\cong\frac{\tau_{M}}{t_{\max}},\qquad\tau_{M}\equiv\intop_{-\infty}^{\infty}d\tau K(\tau),\label{tau_M_def}
\end{equation}
where $\tau_{M}$ is melting time. If there is a true SG transition
on temperature, then melting time should diverge on approaching it
from above. Studying the coefficient in the asymptotic $1/t_{\max}$
form of the glassy order parameter $q$ above freezing could yield
the value of the freezing temperature.

One can also compute the autocorrelation function of the average spin
(magnetization)
\begin{equation}
\mathbf{m}=\frac{1}{N}\sum_{i}\mathbf{s}_{i}\label{m_def}
\end{equation}
that is defined by
\begin{equation}
C(\tau)=\mathbf{m}(t)\cdot\mathbf{m}(t+\tau),\label{C_def}
\end{equation}
In simulations on finite-size systems $C(\tau)$ is non-zero and can
be used to monitor freezing. Unlike $K(\tau)$, it can be computed
for large systems and large time intervals.

The value of the equal-time correlation function $C(0)=m^{2}$ is
nonzero in finite-size systems even in the absence of ling-range order
due to short-range correlations. One has
\begin{equation}
m^{2}=\frac{1}{N^{2}}\sum_{i,j}\mathbf{s}_{i}\cdot{\bf s}_{j}=\frac{1}{N}\sum_{j}\langle{\bf s}_{i}\cdot{\bf s}_{i+j}\rangle\Rightarrow\frac{1}{N}\intop_{0}^{\infty}\frac{d^{d}r}{a^{d}}G(r),
\end{equation}
where $G(r)$ is the spatial correlation function and $d$ is the
dimensionality of the space. As the RA magnet has lots of metastable
local energy minima, $G(r)$ depends on the initial conditions and
on the details of the energy minimization routine. In 2D for $G(r)=\exp\left[-\left(r/R_{f}\right)^{p}\right]$
one obtains
\begin{equation}
m^{2}=K_{p}\frac{\pi R_{f}^{2}}{Na^{2}}\quad\Longrightarrow\quad\frac{R_{f}}{a}=m\sqrt{\frac{N}{\pi K_{p}}},\label{Rf_via_m}
\end{equation}
where $K_{1}=2$ and $K_{2}=1$.

Having estimated $R_{f}$, one can find the number of IM domains $N_{IM}$
in the system of size used in the numerical work. In 2D with linear
sizes $L_{x}$ and $L_{y}$ one has $N_{IM}=L_{x}L_{y}/\left(\pi R_{f}^{2}\right)$.
In particular, for a system with $N=300\times340=102000$ spins and
$D_{R}/J=0.3$, energy minimization at $T=0$ starting from a random
spin state yields $m\approx0.2$1, and with $p=2$ one obtains $R_{f}/a\approx37.8$
and $N_{IM}\approx23$. For $D_{R}/J=1$, one obtains $m\approx0.074$
and $R_{f}/a\approx13.3$ that yields $N_{IM}\approx183$. For the
ratio of the $R_{f}$ values one obtains $R_{f}^{(D_{R}=0.3)}/R_{f}^{(D_{R}=1)}\approx2.84$
that is close to the value 3.33 given by Eq. (\ref{Rf}).

One can compute the linear static susceptibility differentiating the
statistical expression for the average magnetization value $\left\langle \mathbf{m}\right\rangle $.
The differential susceptibility per spin has the form
\begin{equation}
\chi_{\alpha\alpha}=\frac{\partial\left\langle m_{\alpha}\right\rangle }{\partial H_{\alpha}}=\frac{N}{T}\left(\left\langle m_{\alpha}^{2}\right\rangle -\left\langle m_{\alpha}\right\rangle ^{2}\right),\label{chi_comps}
\end{equation}
where the average is taken over the statistical ensemble and $\alpha=x,y,z$.
Within the Monte Carlo method, the average is taken over the statistical
sample generated by the Monte Carlo process. The symmetrized form
of the susceptibility in zero field is given by
\begin{equation}
\chi=\frac{N}{3T}\left(\left\langle \mathbf{m}\cdot\mathbf{m}\right\rangle -\left\langle \mathbf{m}\right\rangle \cdot\left\langle \mathbf{m}\right\rangle \right),\label{chi_symm}
\end{equation}
where $\left\langle \mathbf{m}\cdot\mathbf{m}\right\rangle =\left\langle m^{2}\right\rangle $.
Whereas the number of spins $N$ is very large, the difference of
the terms in brackets can be very small below freezing, so that the
result for a large system does not essentially depend on $N$. Unlike
the magnetization value $\left\langle \mathbf{m}\right\rangle $ that
for a large system can be computed using only one system's state,
i.e., without averaging, $\left\langle \mathbf{m}\right\rangle \Rightarrow\mathbf{m}$,
computing $\chi$ requires averaging over different states of the
statistical ensemble. With only one state taken into account, $\chi$
vanishes. Above the freezing temperature in zero field, one has $\left\langle \mathbf{m}\right\rangle =0$,
so that only the first term in the susceptibility formula contributes.
In the frozen state, the two terms are close to each other and their
difference is small. For this reason, there is a lot of numerical
noise in this formula. In the intermediate temperature range the system
does not reach equilibrium during the simulation time, so that Eq.
(\ref{chi_symm}) becomes questionable as it was obtained under the
assumption of equilibrium using the statistical ensemble. The fact
that the system does not come to equilibrium is another source of
the noise in the simulation results for $\chi$. In different simulations,
the system is getting stuck in one of the infinite number of energy
valleys of its phase space that are characterized by different values
of $\left\langle \mathbf{m}\right\rangle $. However, even in this
intermediate region the formula gives plausible results and should
be correct at least qualitatively. Using $\chi_{\alpha\alpha}=\partial\left\langle m_{\alpha}\right\rangle /\partial H_{\alpha}$
is not much better as the result depends on the time allowed for the
system to relax. At high and low temperatures Eq. (\ref{chi_symm})
is correct as either global or local equilibrium is reached.

\section{The numerical method}

\label{sec:The-numerical-method}

\begin{figure}
\begin{centering}
\includegraphics[width=9cm]{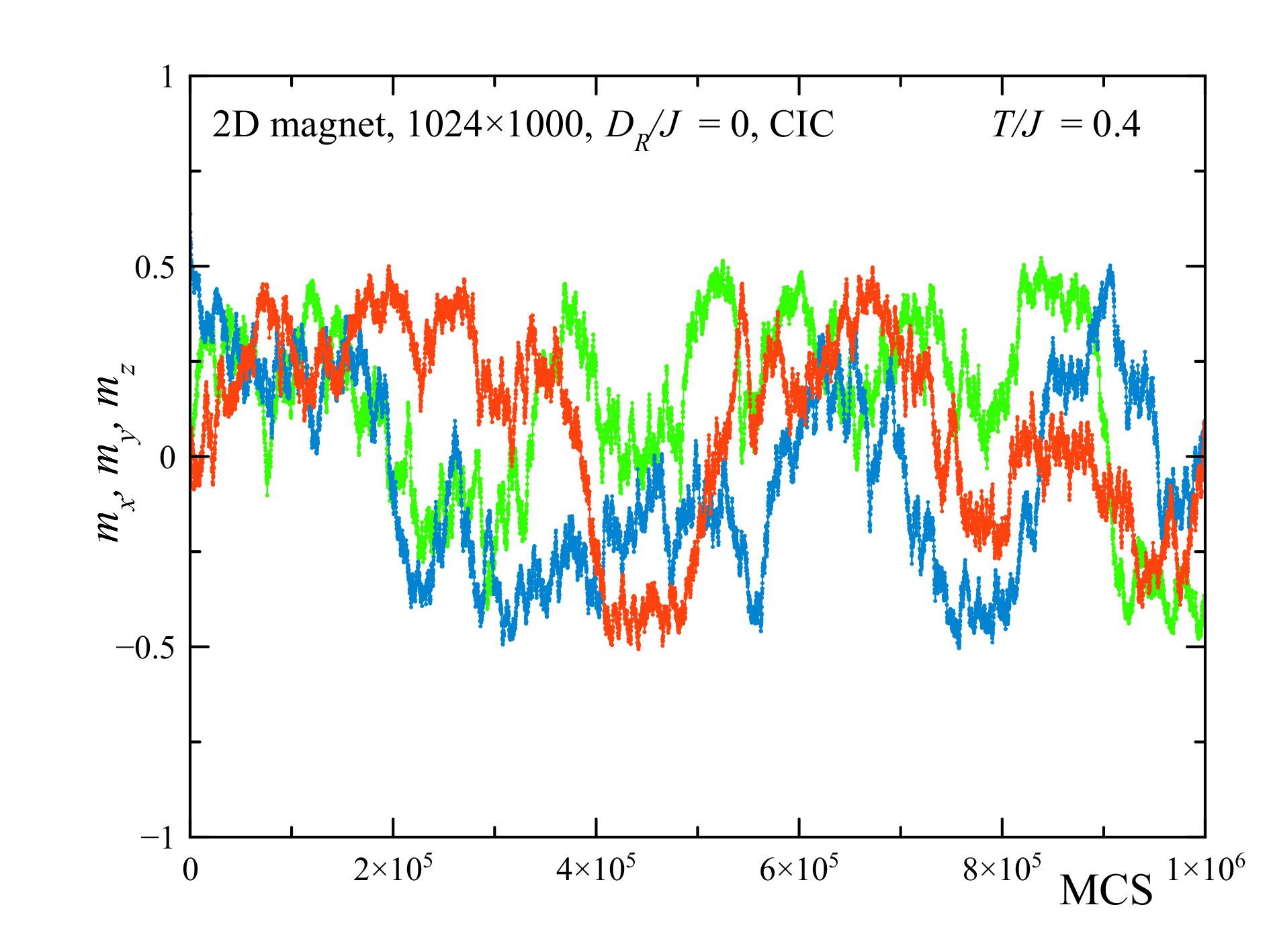}
\par\end{centering}
\begin{centering}
\includegraphics[width=9cm]{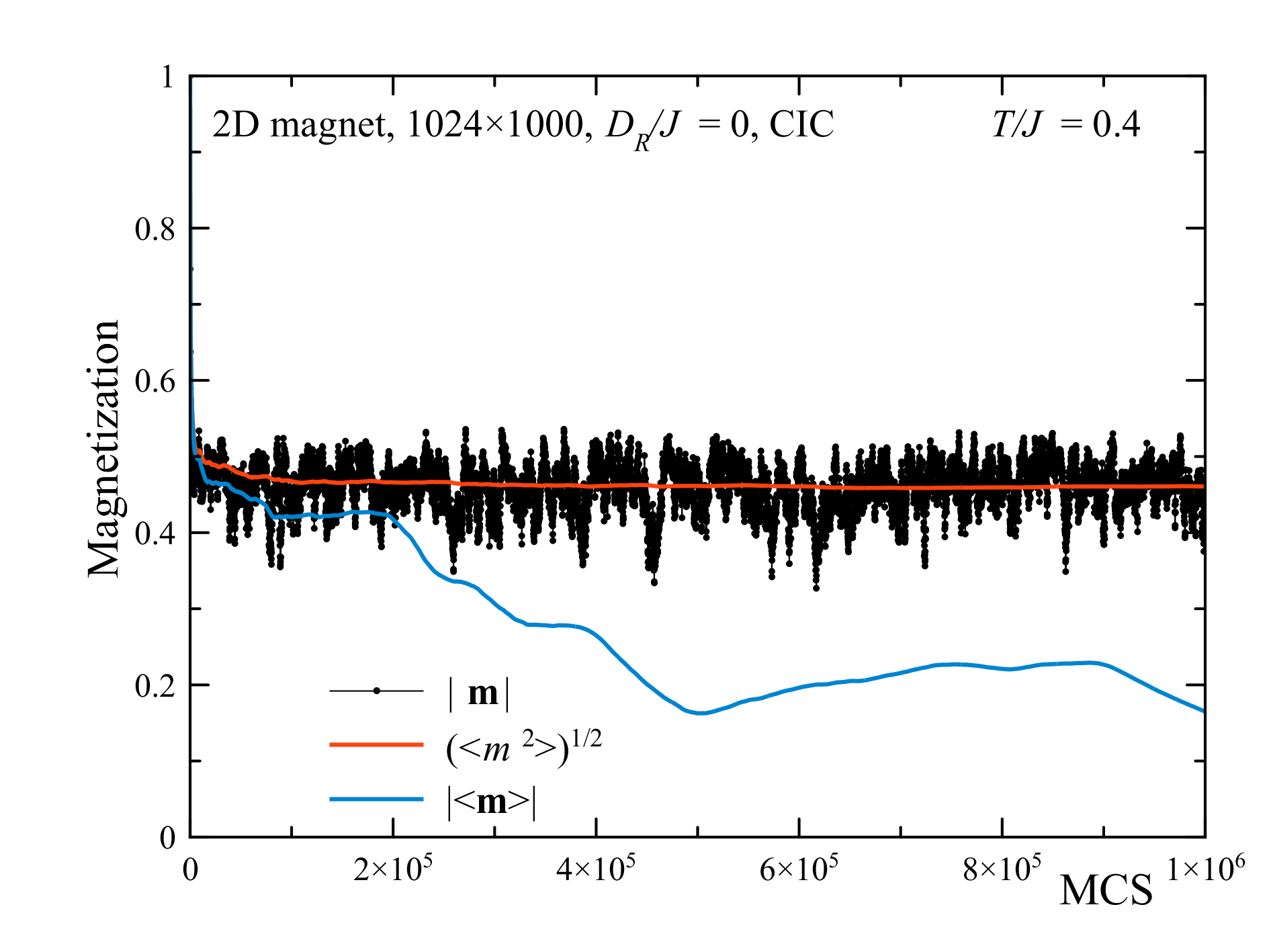}
\par\end{centering}
\caption{Pure 2D Heisenberg ferromagnetic model. Upper panel: Evolution of
the magnetization components. Lower panel: Magnetization, root-mean-square
magnetization, and its running average.}

\label{Fig-mx_my_mz_DR=00003D0}
\end{figure}
\begin{figure}
\begin{centering}
\includegraphics[width=9cm]{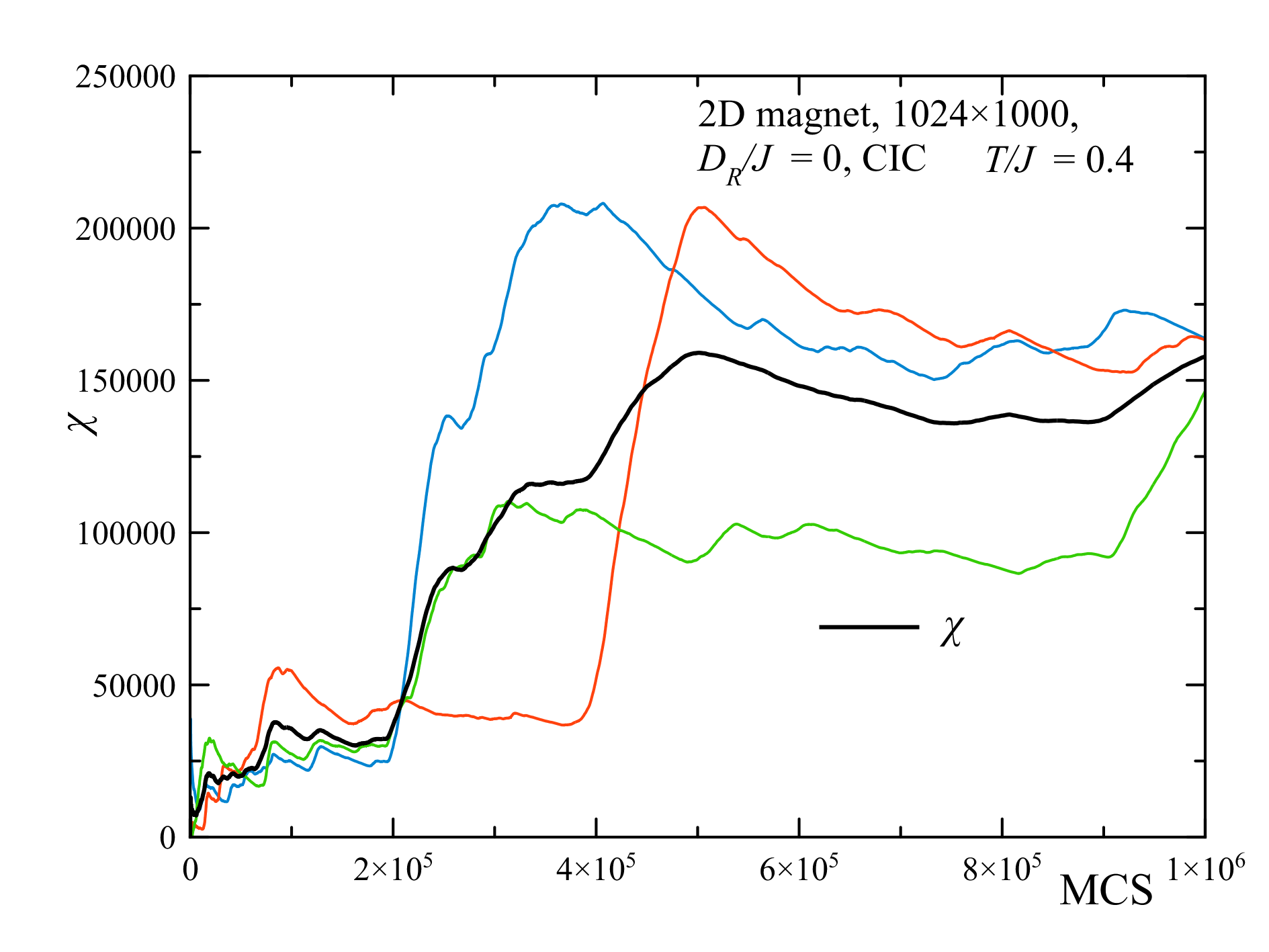}\caption{Susceptibility vs simulation time in MCS for the pure 2D Heisenberg
ferromagnetic model.}
\par\end{centering}
\label{Fig-chix_chiy_chiz_DR=00003D0}
\end{figure}

\begin{figure}
\begin{centering}
\includegraphics[width=9cm]{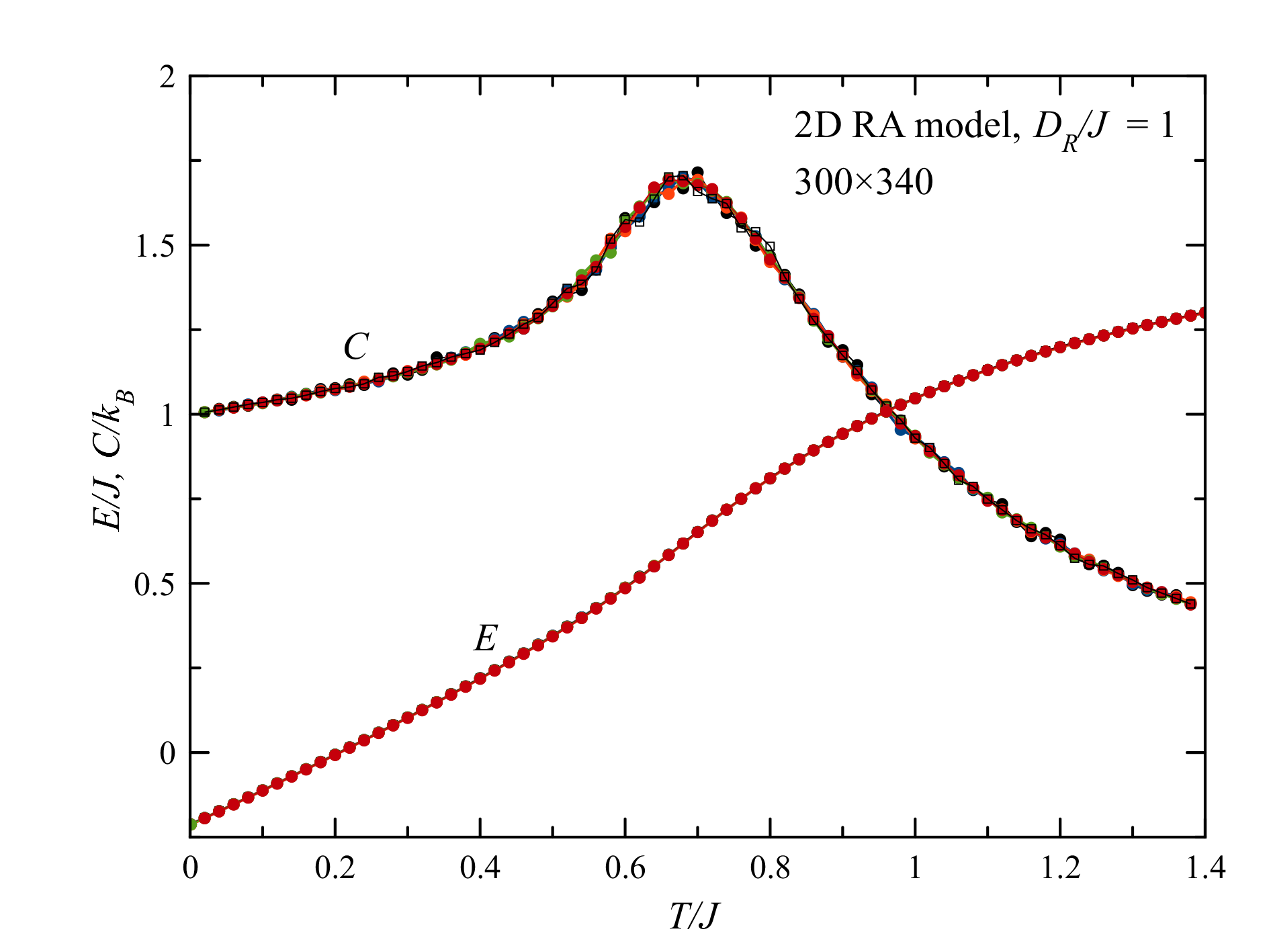}
\par\end{centering}
\caption{Energy and heat capacity of the 2D RA model.}

\label{Fig-C_vs_T_2D_DR=00003D1}
\end{figure}

\begin{figure}
\begin{centering}
\includegraphics[width=9cm]{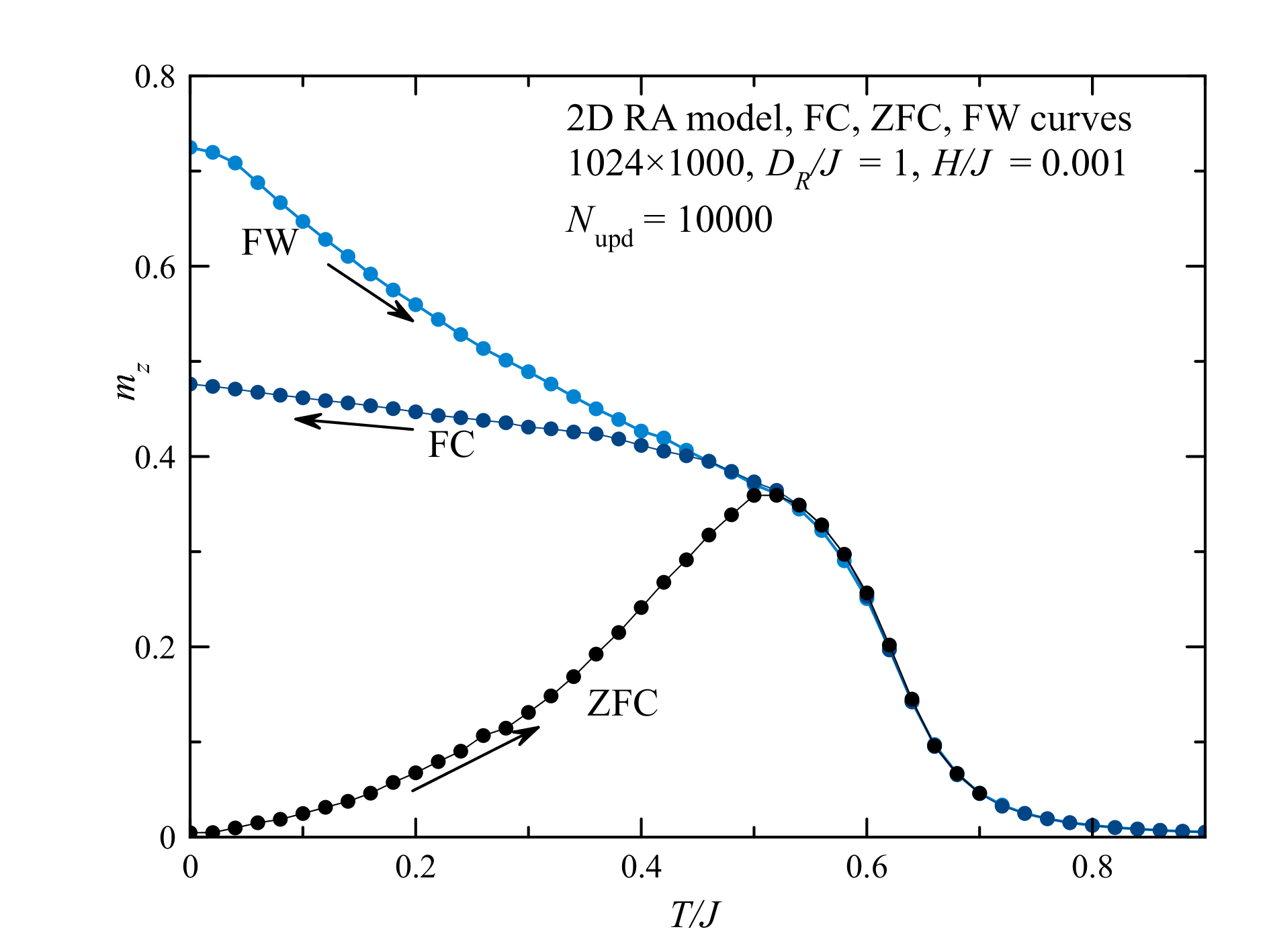}
\par\end{centering}
\caption{FC, ZFC, and FW curves for a 2D RA model with $10^{6}$ spins.}

\label{Fig-FC_ZFC_FW}
\end{figure}

Computation of the static properties of RA magnets at $T>0$ could
be done using real dynamics \citep{garchu_arXiv} or Monte Carlo.
The latter is much faster and is the way to go. However, as different
states of the RA magnet are separated by energy barriers and spins
are fluctuating as large correlated groups, there is a very slow relaxation
near and below the freezing point, creating a computational challenge.
To speed-up the relaxation in simulations, one has to combine the
Metropolis Monte Carlo updates that work as slow diffusion in the
phase space of the system with overrelaxation updates that simulate
conservative dynamics allowing to quickly explore the hypersurfaces
of constant energy. For systems with single-site anisotropy, the straightforward
overrelaxation routine, rotating the spins by $180\lyxmathsym{\textdegree}$
around the effective field, leads to the energy decrease and thus
is not working properly.

To beat this problem, we have developed a \textit{thermalized overrelaxation}
routine. Here, the spins are rotated by $180\lyxmathsym{\textdegree}$
around the different-site part of the effective field (e.g., around
the exchange field). As the result, the energy increases or decreases
due to the RA. To compensate for this, the rotation is accepted or
rejected using the Metropolis criterion, same as in the Monte Carlo
updates. In our simulations, for each spin update we used the Metropolis
Monte Carlo with the probability $\alpha=0.1$ and thermalized overrelaxation
with the probability $1-\alpha$. The thermodynamic consistency of
this method has been checked by computing the dynamical spin temperature
$T_{S}$ given by Eq. (9) of Ref. \citep{Garanin}. The values of
$T_{S}$ were in a good accordance with the set temperature $T$.

To minimize the energy of the system at $T=0$, we used the straightforward
overrelaxation routine mentioned above that for the systems with uniaxial
single-site anisotropy leads to the energy decrease. This routine
provides a fast convergence. If in the initial state of the system
all spins are collinear, then upon relaxation the system becomes only
partially disordered with a significant residual magnetization $m$.
If the initial state of the system is random, the relaxed state is
random, too, with $m$ being rather small. Most of the simulations
were done with the random initial conditions (RIC).

In the computation of the dependence of the SG order parameter $q$
on $t_{\max}$, Eq. (\ref{q_def}), we used summation over Monte Carlo
steps (MCS) i.e., system updates, instead of the integration over
time. That is, instead of Eq. (\ref{s_avr_t}) we used
\begin{equation}
\left\langle \mathbf{s}_{i}\right\rangle \equiv\frac{1}{\mathrm{MCS}}\sum_{n=1}^{\mathrm{MCS}}\mathbf{s}_{i}(n),\label{s_avr_MCS}
\end{equation}
where $n$ labels the states generated by the Monte Carlo process.
The asymptotic formula for $q$ above freezing becomes
\begin{equation}
q\cong\frac{\tau_{M}}{\mathrm{MCS}},\qquad\tau_{M}=\sum_{n=-\infty}^{\infty}K(n).\label{tau_M_MCS}
\end{equation}
Here, the melting time $\tau_{M}$ is measured in Monte Carlo steps.
While the latter are not related to the time in any simple way, still
$\tau_{M}$ gives an idea if freezing in the system.

The simulations were done on large 2D systems, typically $1024\times1000$
spins, with periodic boundary conditions. The very large system size
is needed as the system of many spins behaves as that of a much smaller
number of IM domains. In particular, for $D_{R}/J=1$ the system of
$10^{5}$ spins is too small and shows large fluctuations. To reduce
fluctuations, one can either perform repeated measurements on such
systems or simulate a larger system. The latter is preferred.

Usually, in Monte Carlo simulations the system is first equilibrated
and then measurements of the equilibrium properties are performed.
For the RA magnet, the equilibration is extremely long, and around
the freezing temperature the system does not come to equilibrium at
all after one million MCS that is about our limit for $10^{6}$ spins.
Thus we resorted to performing a fixed number of MCS for each temperature.
Our first attempts included stepwise lowering $T$ performing from
ten to twenty thousand MCS for each temperature point. However, such
simulation duration proved to be too short, and increasing it in the
cycle over the temperatures would result in exceedingly long computation.
Therefore, long simulations, up to $10^{6}$ MCS, for select temperature
values have been performed, each time starting from quenched states
obtained by energy minimization starting from random spin states.
Each of these simulations required several days.

As the computing software, Wolfram Mathematica with compilation and
parallelization was used. Most of computations were performed on our
Dell Precision workstation having 20 CPU cores from which 16 cores
were used by Mathematica.

\section{Testing the numerical method on the pure 2D magnet}

\label{sec:Testing-the-numerical}

First, we test the numerical method on the pure 2D Heisenberg ferromagnetic
model. In this case, the thermalized overrelaxation degenerates to
the regular overrelaxation as it conserves energy. There is no phase
transition on temperature in this model but a strong short-range order
with exponentially large magnetic susceptibility and exponentially
long correlation length establishes with lowering temperature in the
infinite system. This happens around $T/J=0.7$ where the heat capacity
has a maximum. In simulations on finite-size systems, further lowering
the temperature quickly results in the correlation length exceeding
the system size $L$, and the system behaves as an ordered magnetic
particle. In this regime, the susceptibility is not exponentially
large but still huge: $\chi=N\left\langle m^{2}\right\rangle /(3T)$.
The results of a single simulation of the pure 2D Heisenberg ferromagnetic
model at $T/J=0.4$ are shown in Figs. \ref{Fig-mx_my_mz_DR=00003D0}
and \ref{Fig-chix_chiy_chiz_DR=00003D0}. The equilibrium value of
the root-mean-square magnetization $\left\langle m^{2}\right\rangle ^{1/2}$
stabilizes quickly enough with increasing the number of system updates
MCS, as can be seen in the lower panel of Fig. \ref{Fig-mx_my_mz_DR=00003D0}).
However, computing the linear susceptibility $\chi$ using Eq. (\ref{chi_symm})
for the system of one million spins requires about a million of system
updates, see Fig. \ref{Fig-chix_chiy_chiz_DR=00003D0}. Such a long
simulation is required to average out the system's magnetization vector
$\mathbf{m}$ that contributes to the second term in Eq. (\ref{chi_symm}).
The slow evolution of the components of $\mathbf{m}$ is shown in
the upper panel of Fig. \ref{Fig-mx_my_mz_DR=00003D0}. After one
million of system updates one obtains $\chi J=157838$ that is almost
as large as the ``magnetic-particle'' value defined just above with
$N=1024\times1000$, $\sqrt{\left\langle m^{2}\right\rangle }=0.46$,
and $T/J=0.4$ that is $\chi J=180565$.

One also can check what becomes the freezing parameter $q$ of Eq.
(\ref{q_def}) for the pure system. Theoretically, one expects $q=0$
at any $T>0$. However, for a system of $10^{6}$ spins at $T/J=0.4$
one needs to perform hundreds of thousands MCS to see that the system
is not frozen. The dependence of $q$ on the number of MCS performed
is added to Fig. \ref{Fig_q_lin-lin} below. These simulations of
the pure system show that the problem is computationally involved.
In the presence of random anisotropy it becomes harder because of
even longer relaxation due to thermally-activated barrier crossing
by large groups of correlated spins.

\section{Numerical results}

\label{sec:Numerical-results}

\begin{figure}
\begin{centering}
\includegraphics[width=9cm]{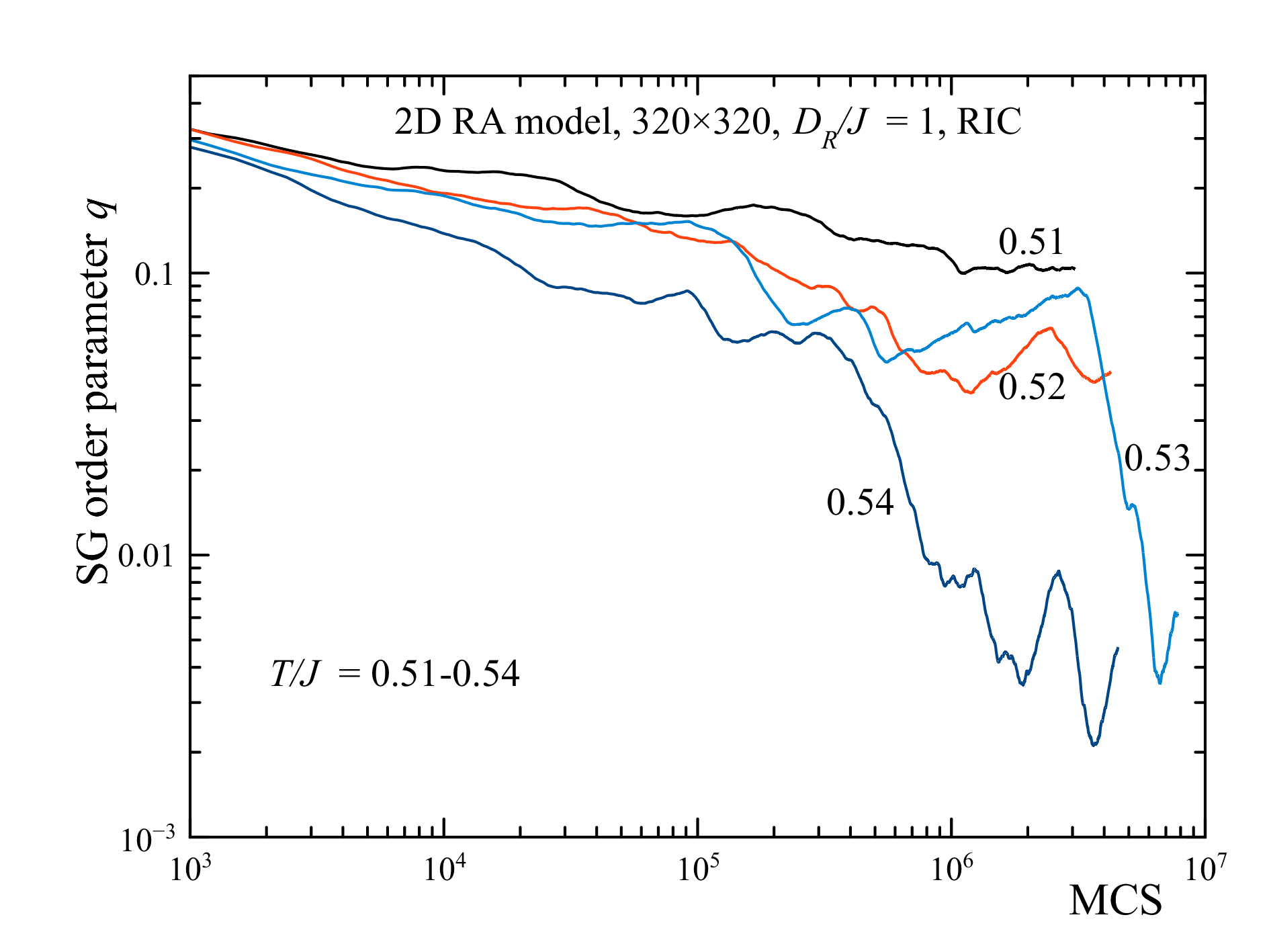}
\par\end{centering}
\begin{centering}
\includegraphics[width=9cm]{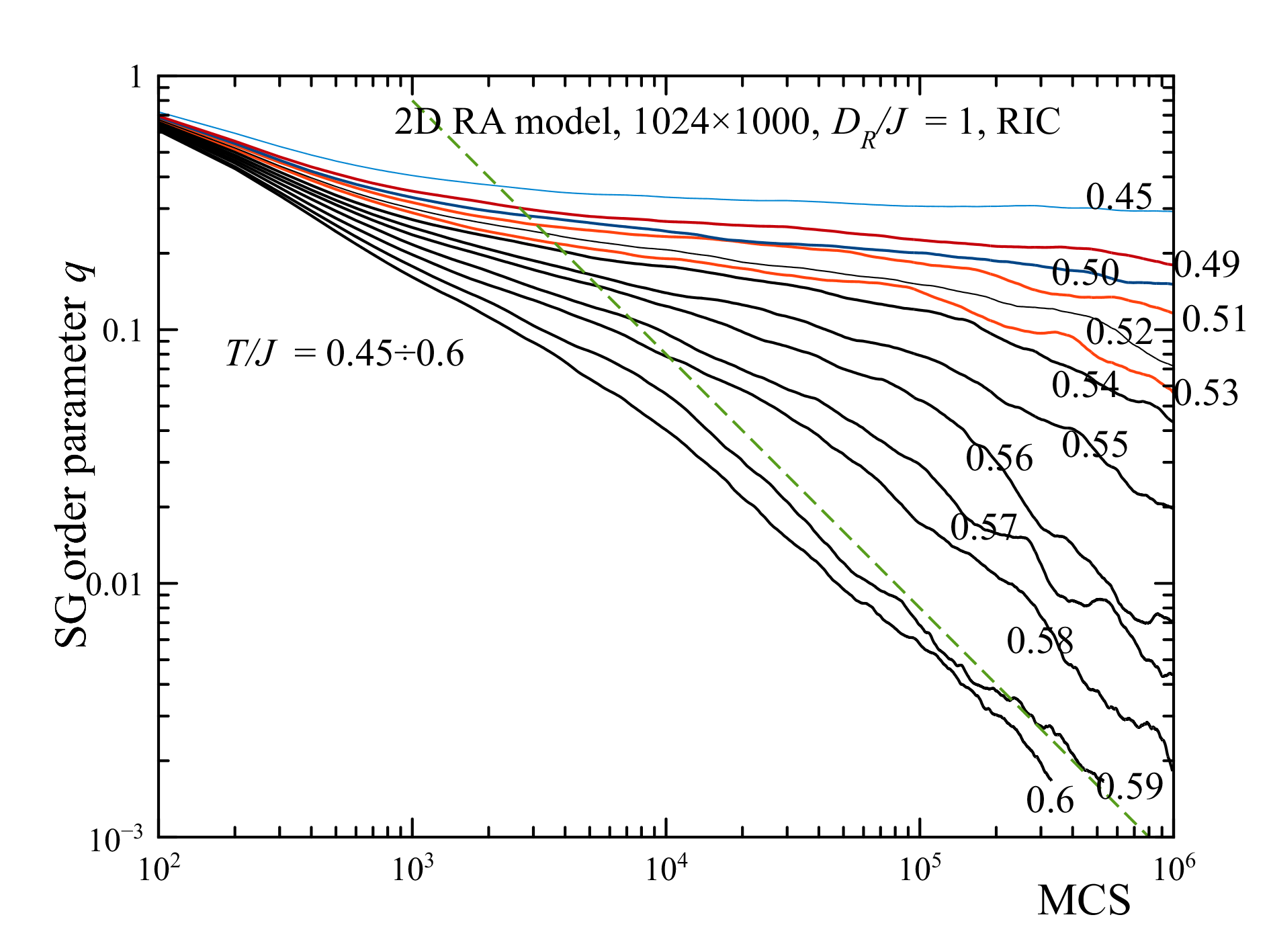}
\par\end{centering}
\caption{The dependence of the SG order parameter on the number of Monte Carlo
steps in log-log scale. Upper panel: a smaller system of $10^{5}$
spins. Lower panel: the system of $10^{6}$ spins. The dashed line
is the asymptote $q=\tau_{M}/\mathrm{MCS}$ for $T/J=0.59$.}

\label{Fig-q2_log-log}
\end{figure}

\begin{figure}
\begin{centering}
\includegraphics[width=9cm]{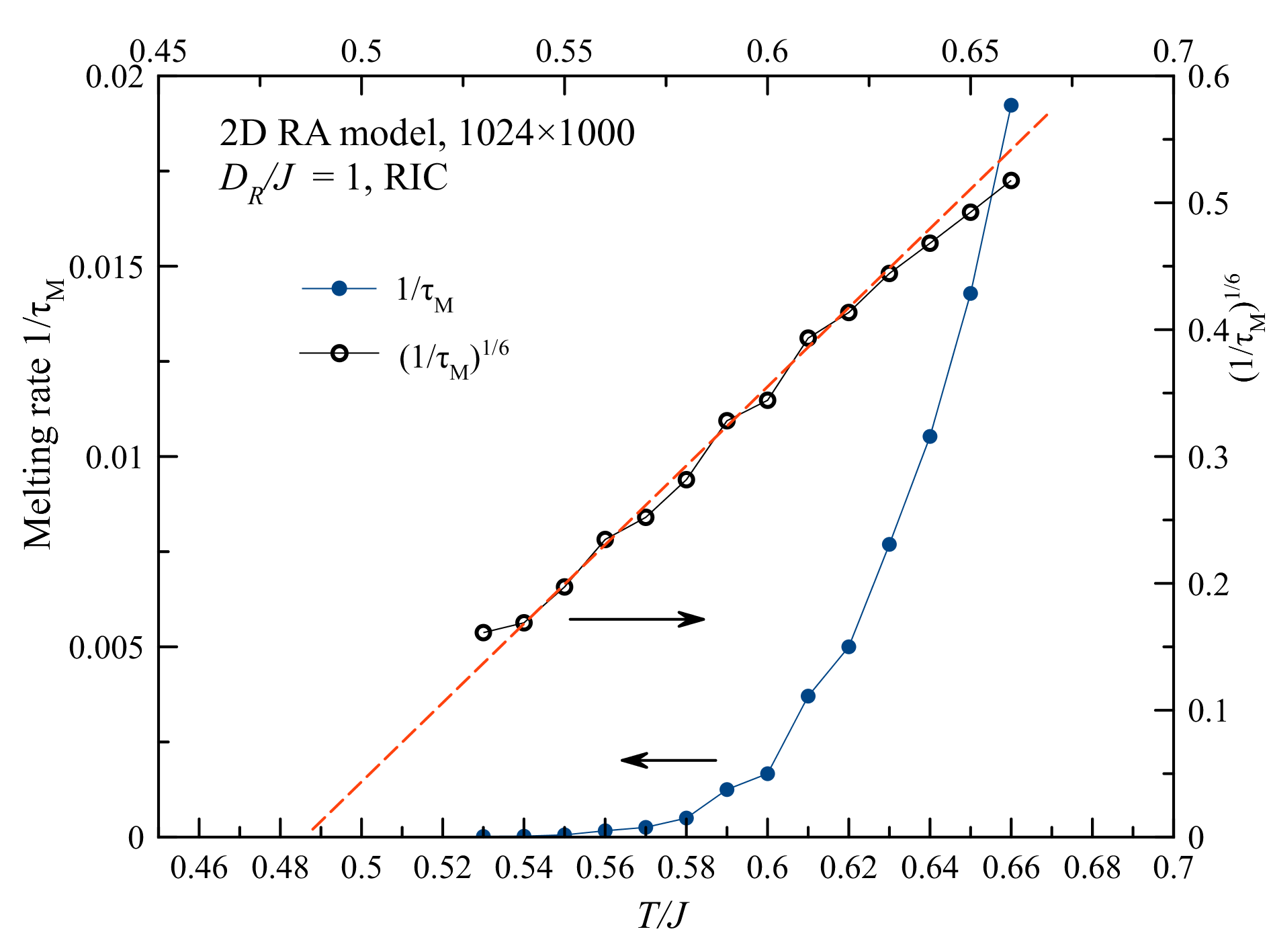}
\par\end{centering}
\caption{Melting rate above freezing -- natural and power-law representations.}

\label{Fig-Melting rate}
\end{figure}
\begin{figure}
\begin{centering}
\includegraphics[width=9cm]{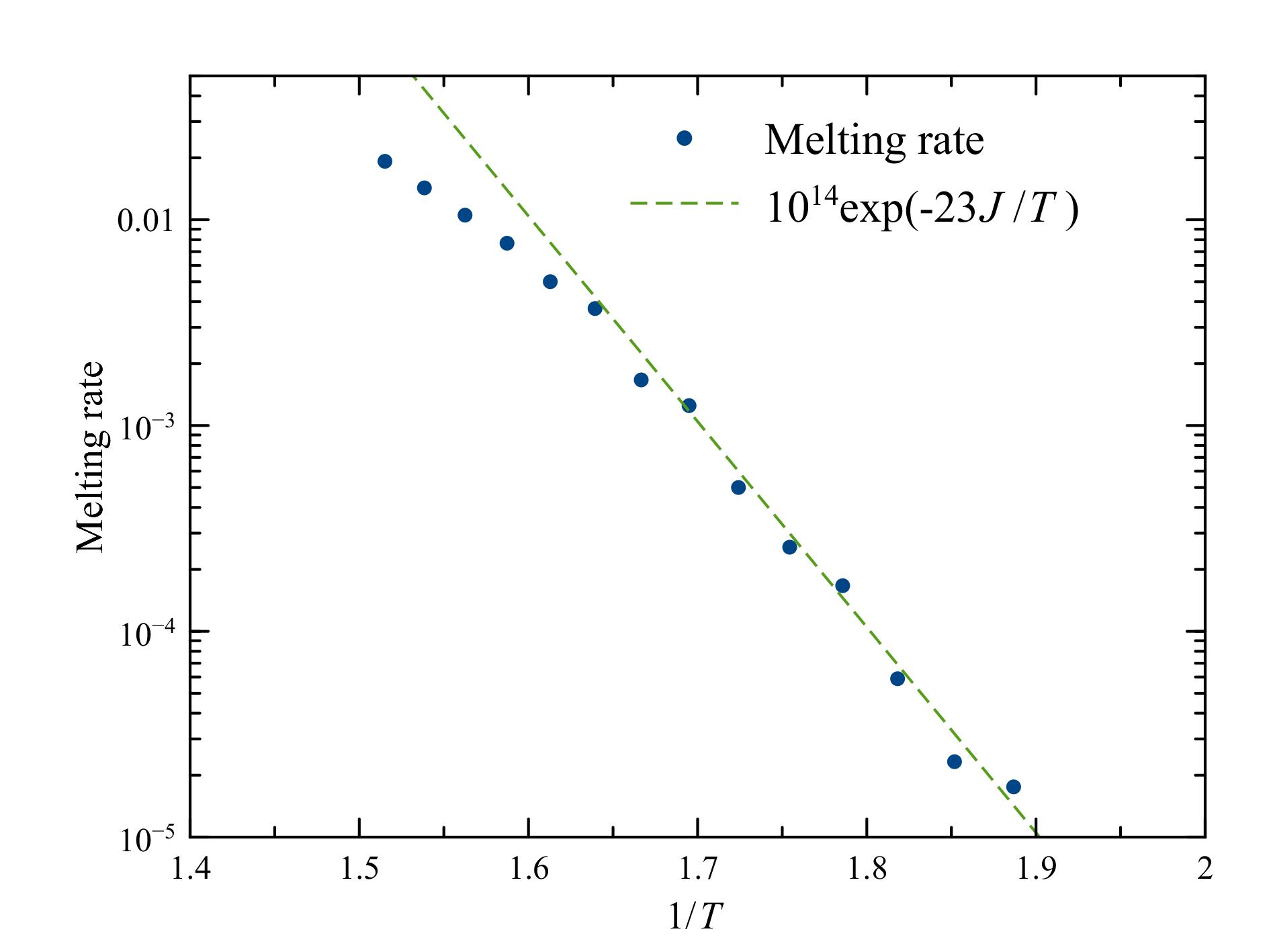}
\par\end{centering}
\caption{Melting rate above freezing -- Arrhenius representation.}

\label{Fig-Melting_rate_Arrhenius}
\end{figure}
\begin{figure}
\begin{centering}
\includegraphics[width=9cm]{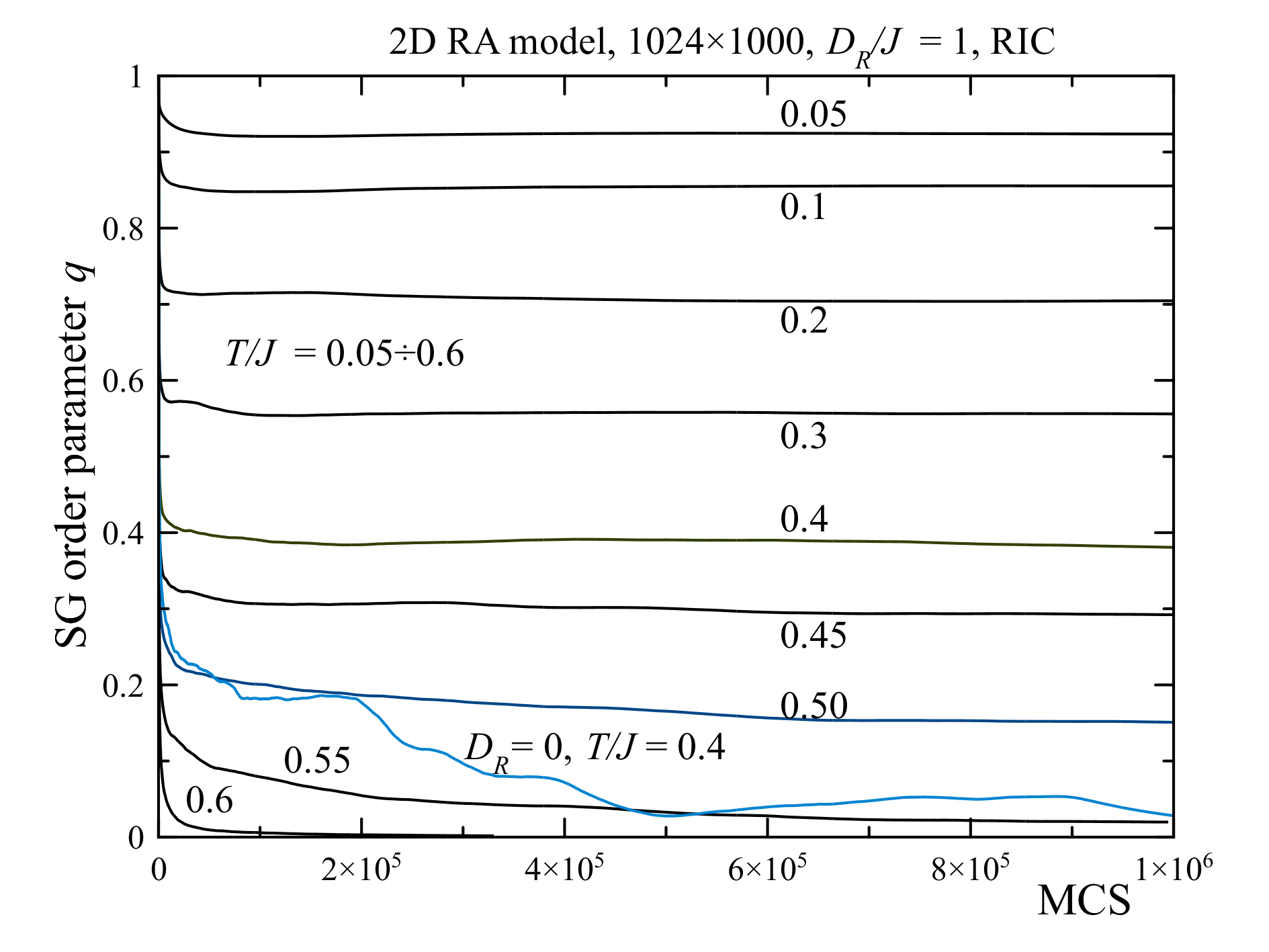}
\par\end{centering}
\caption{The dependence of the SG order parameter on the number of Monte Carlo
steps in linear scale at different temperatures.}

\label{Fig_q_lin-lin}
\end{figure}
\begin{figure}
\begin{centering}
\includegraphics[width=9cm]{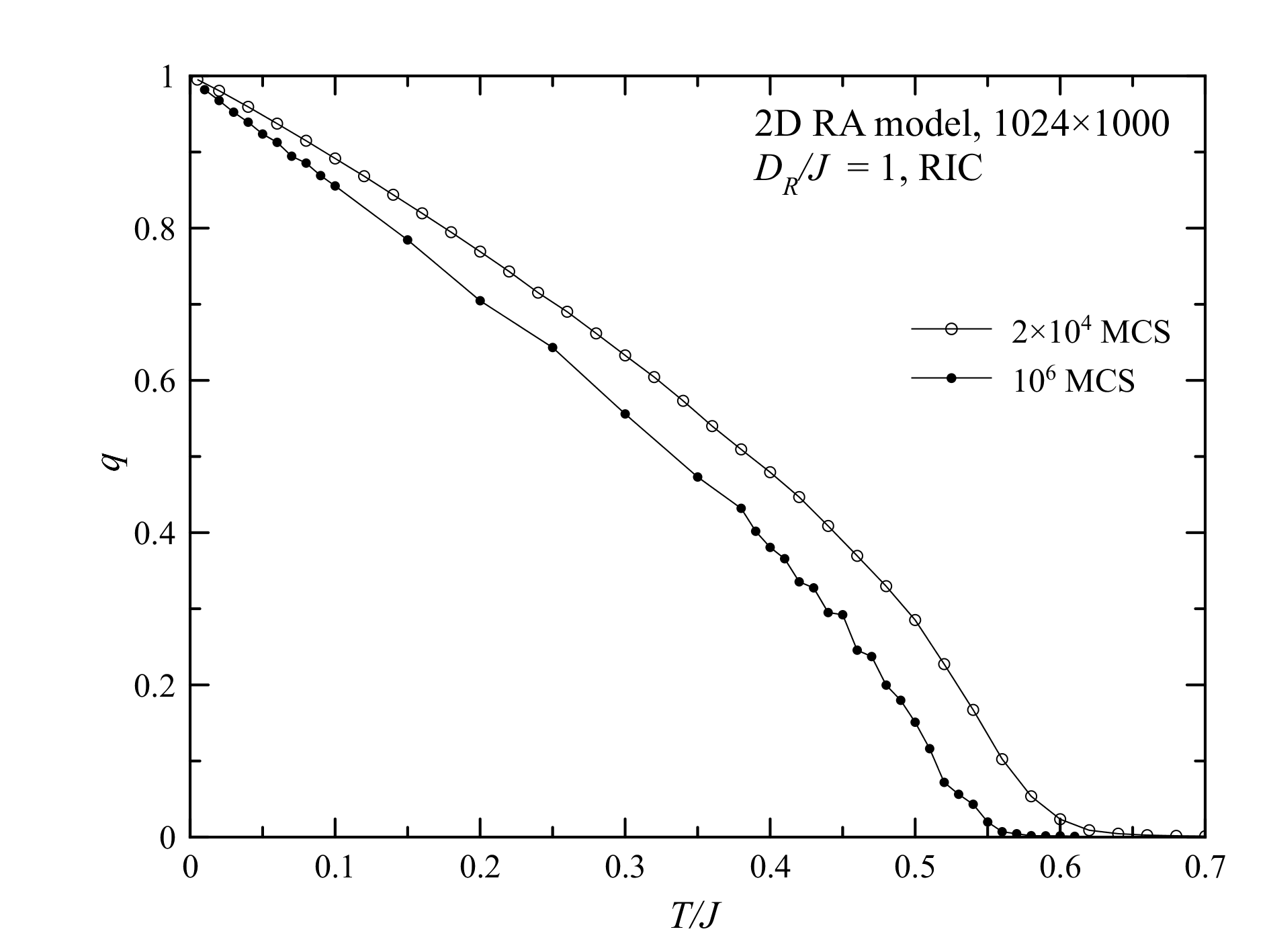}
\par\end{centering}
\caption{Asymptotic values of the spin-glass order parameter $q$ at different
temperatures. The curve with open circles was obtained by gradually
lowering the temperature making $2\times10^{4}$ MCS for each $T$
value. The curve with filled circles was obtained by making $10^{6}$
MCS at each $T$ value one by one for different RA realizations. }

\label{Fig_q_vs_T}
\end{figure}

\subsection{FC-ZFC-FW curves}

Figure \ref{Fig-FC_ZFC_FW} shows the result for the FC, ZFC, and
FW curves in a 2D RA model with the a million spins and the RA strength
$D_{R}/J=1$. Here a very weak field is applied along $z$ axis and
$m_{z}\equiv\left\langle s_{i,z}\right\rangle $ was measured. The
ZFC curve was obtained by first minimizing the system's energy starting
from a random orientation of spins at $T=0$ and $H=0$, than applying
the field $H$ and gradual warming the system. The FC curve was obtained
by gradual cooling the system from a high temperature to $T=0$ in
the presence of the field. Finally, the field-warmed (FW) curve was
obtained by first minimizing the system's energy in the applied field
starting from the state with all spins directed along $z$ axis and
then gradually warming the system. For each temperature point, 10000
system updates were performed. One can see that the ZFC curve merges
with the other two curves at $T/J\simeq0.53$ that can be interpreted
as spin-glass transition or freezing temperature $T_{SG}$. The estimated
magnetic susceptibility near freezing is huge, $\chi=m_{z}/H\simeq400$.
This is unlike that in the conventional spin glasses with a random
exchange. Each curve was obtained by averaging over three different
runs. For a system of $10^{5}$ spins fluctuations are much stronger,
so that a more extensive averaging over runs is needed.

\subsection{Spin-glass order parameter and melting time}

Then, we have performed a long annealing, up to $10^{6}$ MCS, of
the system at different temperatures after the energy minimization
(quenching) starting from random initial conditions. Each of these
simulations took several days, so they had to be done one-by-one rather
than in a cycle. Each simulation run used its own realization of the
RA and its own random initial spin state. The results of each run
were the dependences of $q$ and the components of the average spin
$\mathbf{m}$, Eq. (\ref{m_def}), vs the number of MCS. From the
$\mathbf{m}$ data, the susceptibility components, Eq. (\ref{chi_comps})
and the symmetrized susceptibility, Eq. (\ref{chi_symm}) were derived.

\begin{figure}
\begin{centering}
\includegraphics[width=9cm]{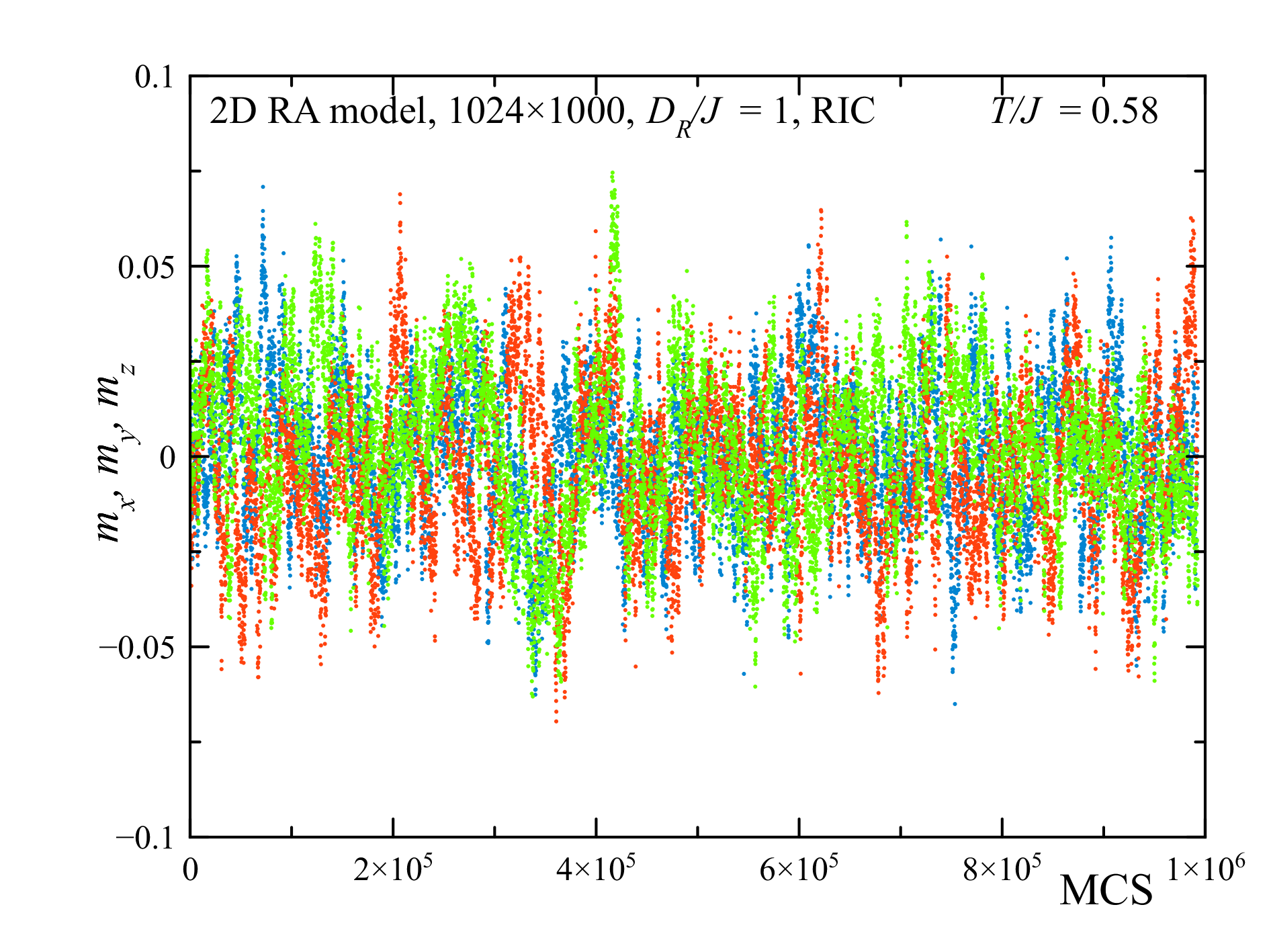}
\par\end{centering}
\begin{centering}
\includegraphics[width=9cm]{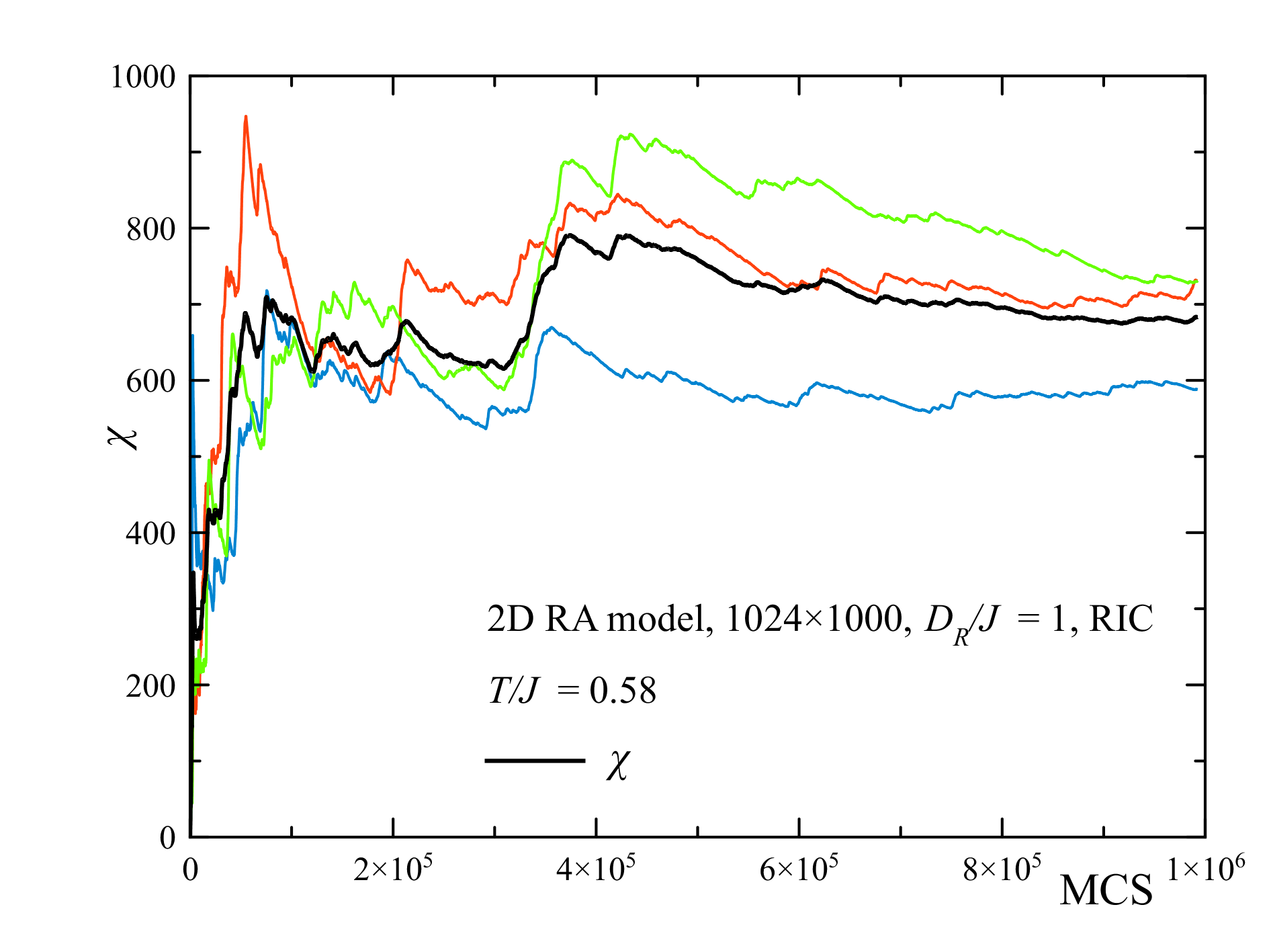}
\par\end{centering}
\caption{Simulation above the freezing point, $T/J=0.58$. Upper panel: evolution
of the magnetization components. Lower panel: dependence of the susceptibility
components (colored curves) and the symmetrized susceptibility (black
curve) on the number of MCS done.}

\label{Fig_T=00003D0.58}
\end{figure}
\begin{figure}
\begin{centering}
\includegraphics[width=9cm]{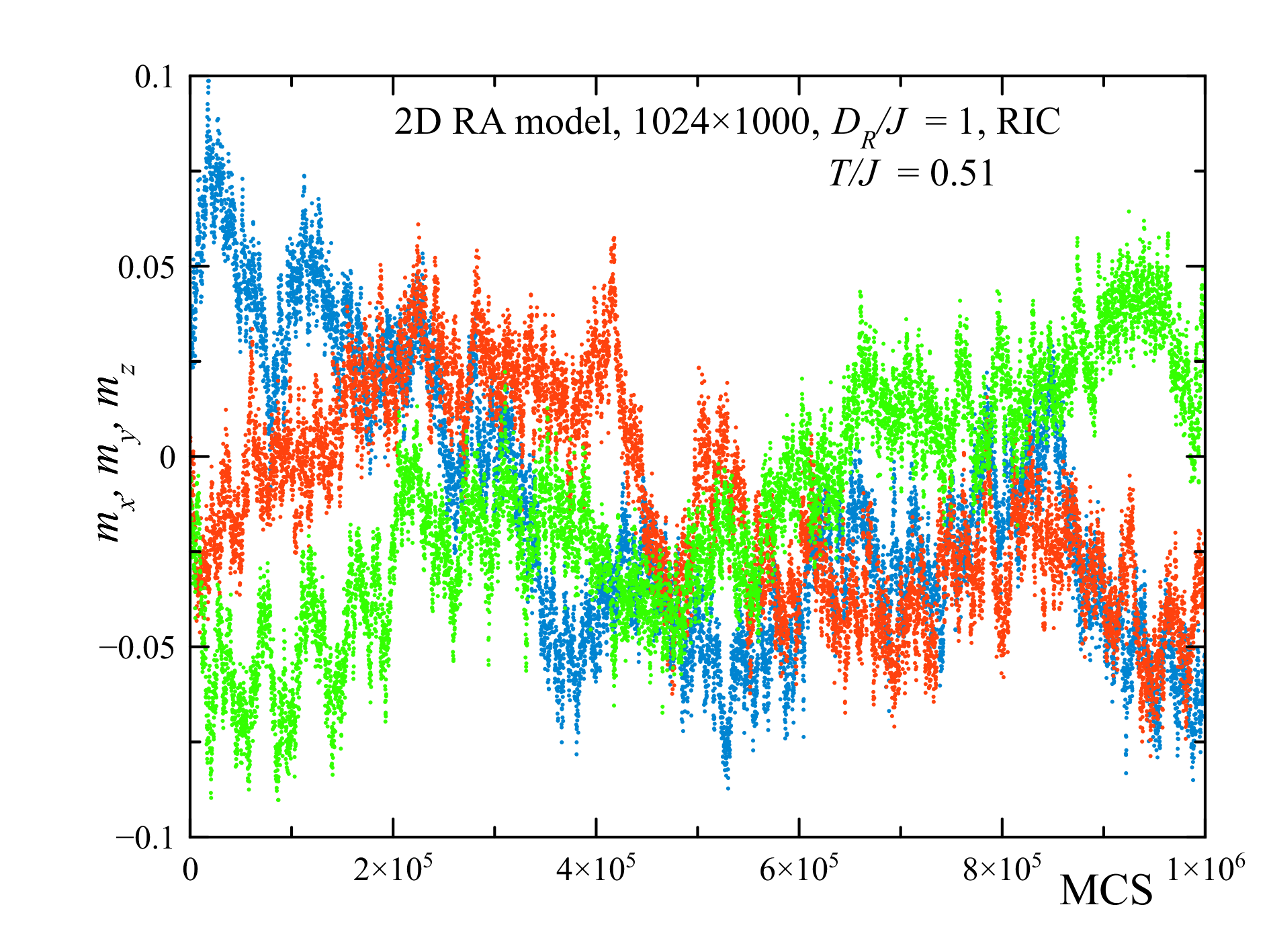}
\par\end{centering}
\begin{centering}
\includegraphics[width=9cm]{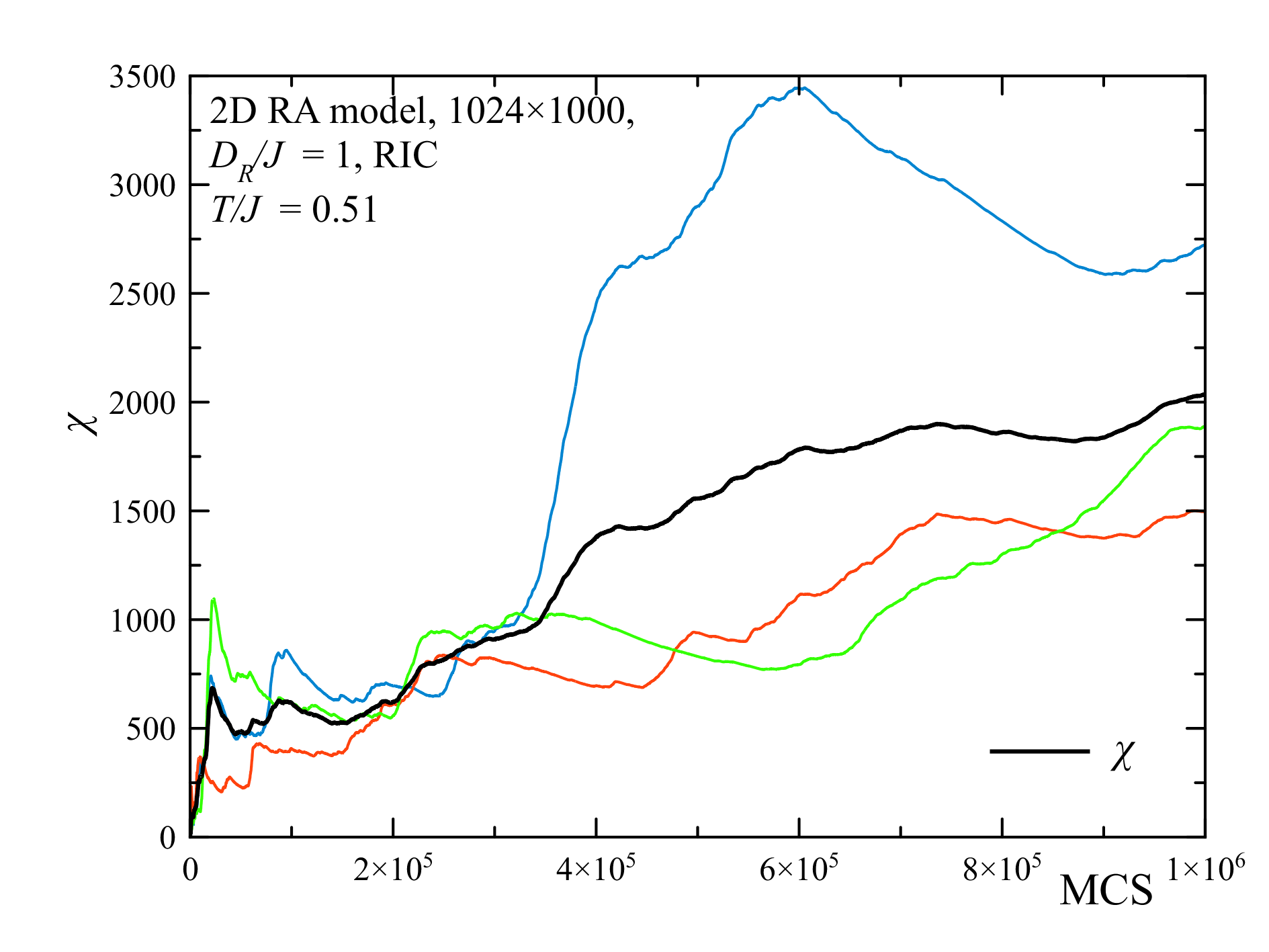}
\par\end{centering}
\caption{Simulation just below the freezing point, $T/J=0.51$. Upper panel:
evolution of the magnetization components. Lower panel: dependence
of the susceptibility components (colored curves) and the symmetrized
susceptibility (black curve) on the number of MCS done. The system
does not come to equilibrium and $\chi$ does not stabilize with increasing
of the number of MCS done.}

\label{Fig_T=00003D0.51}
\end{figure}
\begin{figure}
\begin{centering}
\includegraphics[width=9cm]{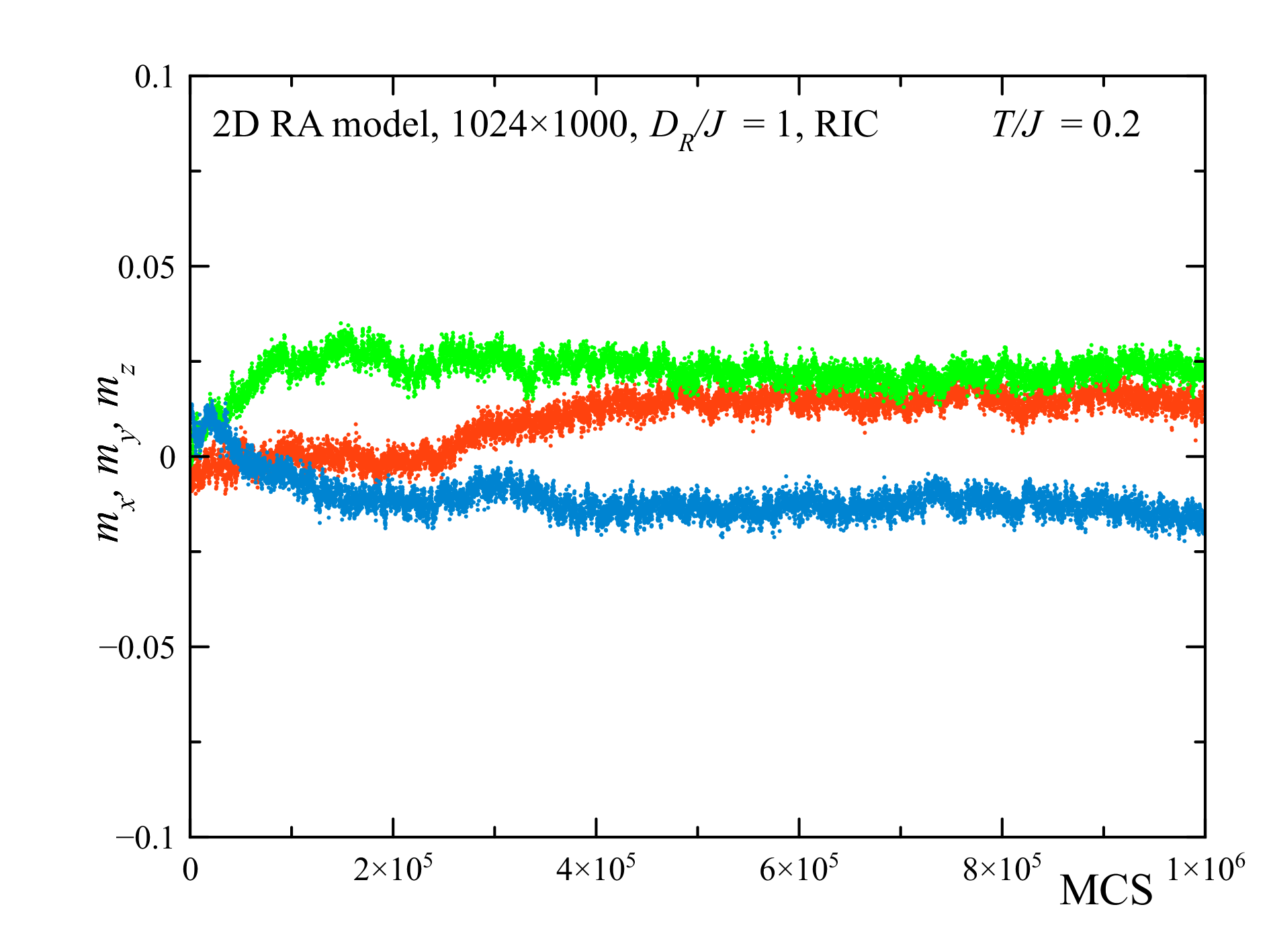}
\par\end{centering}
\begin{centering}
\includegraphics[width=9cm]{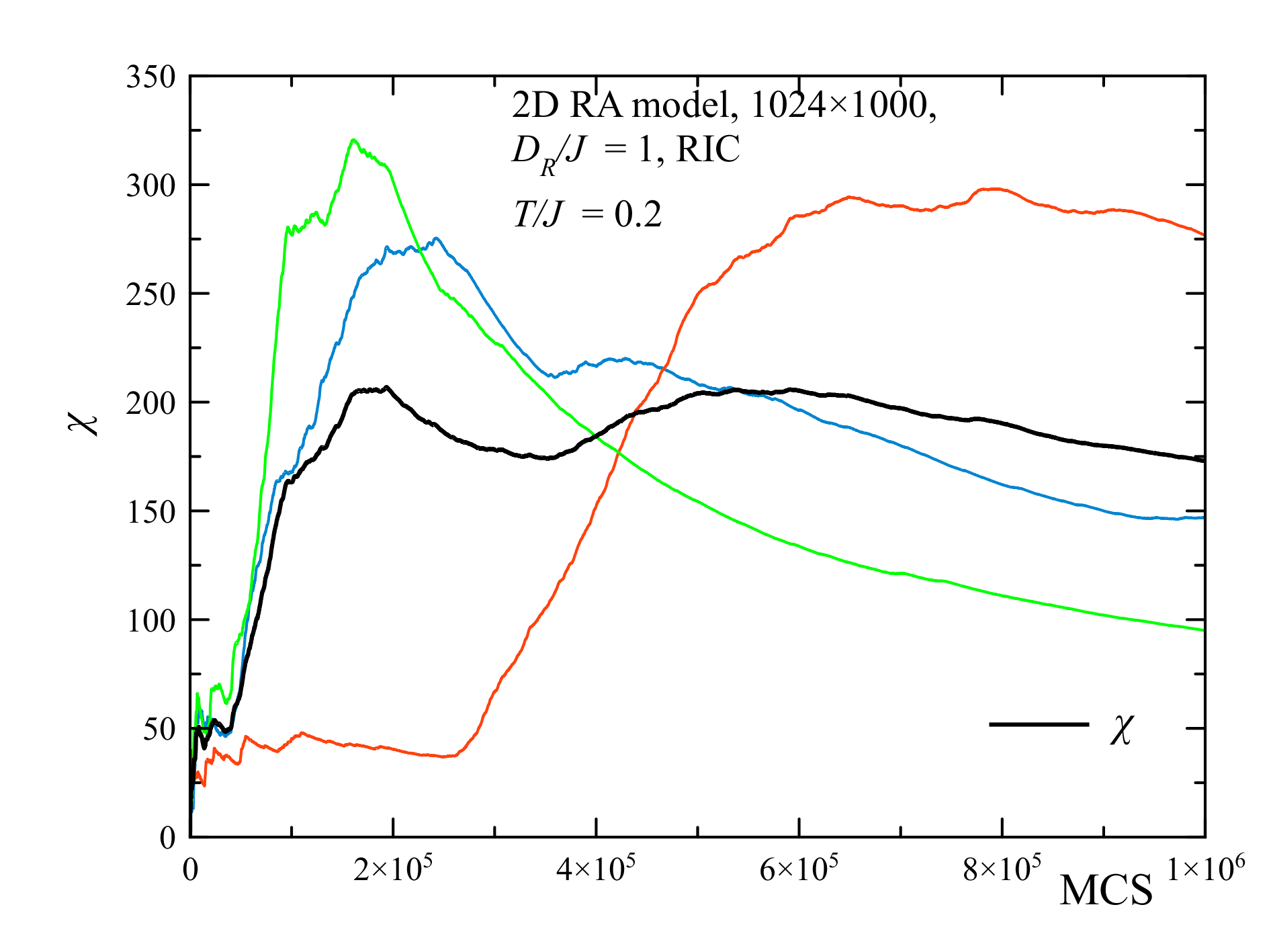}
\par\end{centering}
\caption{Simulation at a lower temperature, $T/J=0.2$. Upper panel: evolution
of the magnetization components; Lower panel: dependence of the susceptibility
components (colored curves) and the symmetrized susceptibility (black
curve) on the number of MCS done.}

\label{Fig_T=00003D0.3}
\end{figure}

The dependence of the SG order parameter $q$ of Eq. (\ref{q_def})
on the number of MCS near the freezing temperature for a smaller system
of $10^{5}$ spins is shown in the upper panel of Fig. \ref{Fig-q2_log-log}.
One can see that this size is too small, as the system behaves as
that of a much fewer number of entities and fluctuations are too strong.
For $T/J=0.53$ at $\mathrm{MCS}=3\times10^{6}$ most of the system's
$10^{5}$ spins suddenly change their direction that leads to a sharp
decrease of $q$. A really large system should not behave like this.
One has either to perform an extensive averaging over runs of take
a larger system that is preferable. The data for $10^{6}$ spins in
the lower panel of Fig. \ref{Fig-q2_log-log} are much smoother and
show the asymptotic power-law dependence $q=\tau_{M}/\mathrm{MCS}$
above the freezing point, in accordance with Eq. (\ref{tau_M_def}).

Fitting the dependence of the SG order parameter with $q=\tau_{M}/\mathrm{MCS}$
one can extract the melting time $\tau_{M}$. The latter becomes very
large when the system freezes. If there is true phase transition at
some freezing temperature $T_{f}$, one can expect a power-law divergence
$\tau_{M}\propto\left(T-T_{f}\right)^{-\gamma}$. The results for
the melting rate $1/\tau_{M}$ are shown in Fig. \ref{Fig-Melting rate}.
The temperature dependence of $1/\tau_{M}^{1/6}$ is a straight line
that suggests $\gamma=6$ and $T_{f}/J\simeq0.49$. The power 6 is
too high to be credible while the freezing temperature is rather low
and difficult to approach from above because of too slow relaxation
requiring exorbitant computing times.

Another way to fit the results for the melting time is using the Arrhenius
temperature dependence $\tau_{M}\propto\exp\left(\Delta U/T\right)$.
The corresponding data representation shown in Fig. \ref{Fig-Melting_rate_Arrhenius}
yields the barrier value $\Delta U=23J$. Theoretically, Eq. (\ref{DeltaU_cases})
yields $\Delta U\sim J$ for $d=2$ but there can be a large numerical
factor in $\Delta U$. At larger temperatures, one can see expected
deviations from the Arrhenius law (as well as deviations from the
power law in Fig, \ref{Fig-Melting rate}). This interpretation implies
that there is no phase transition and freezing is a gradual process.

Another argument in favor of a gradual freezing/melting is the fact
that in systems with quenched disorder the properties averaged over
large regions should fluctuate. If in one region the averaged random
anisotropy is larger than in the others, freezing /melting in this
region will occur at slightly higher temperatures. Thus the freezing
temperature will be spread. At some temperature, most of the system
will be melted while some minoruty regions will be still frozen, providing
a small but nonzero value of the spin-glass order parameter $q$.
In this scenario, $q(T)$ dependence is neither a power nor an exponential
of the temperature.

In the frozen state, as can be seen in Fig. \ref{Fig_q_lin-lin},
the SG order parameter quickly reaches its asymptotic value that is
smaller than one because of the thermal motion of spins on IM domains
in their valleys without crossing the barriers to different valleys.
With increasing the temperature towards the melting point, the processes
of crossing the barriers begin and $q$ slowly decreases. Above the
freezing point, such as $T/J=0.6$, the SG order parameter quickly
decreases to zero.

The values of the SG order parameter $q$ computed at different temperatures
are shown in Fig. \ref{Fig_q_vs_T}. The curve with open circles was
obtained by gradually lowering the temperature making $2\times10^{4}$
MCS for each $T$ value. This duration of annealing is insufficient
to reach stable results. The points of the curve with filled circled
obtained one by one for different RA realizations with $10^{6}$ MCS
are significantly shifted down. In this case, the equilibrium is reached
in the main part of the temperature interval except for the vicinity
of the freezing transition. This is confirmed by the plateaus of $q$
vs the number of MCS in Fig. \ref{Fig_q_lin-lin}.

\subsection{Magnetization and susceptibility}

\begin{figure}
\begin{centering}
\includegraphics[width=9cm]{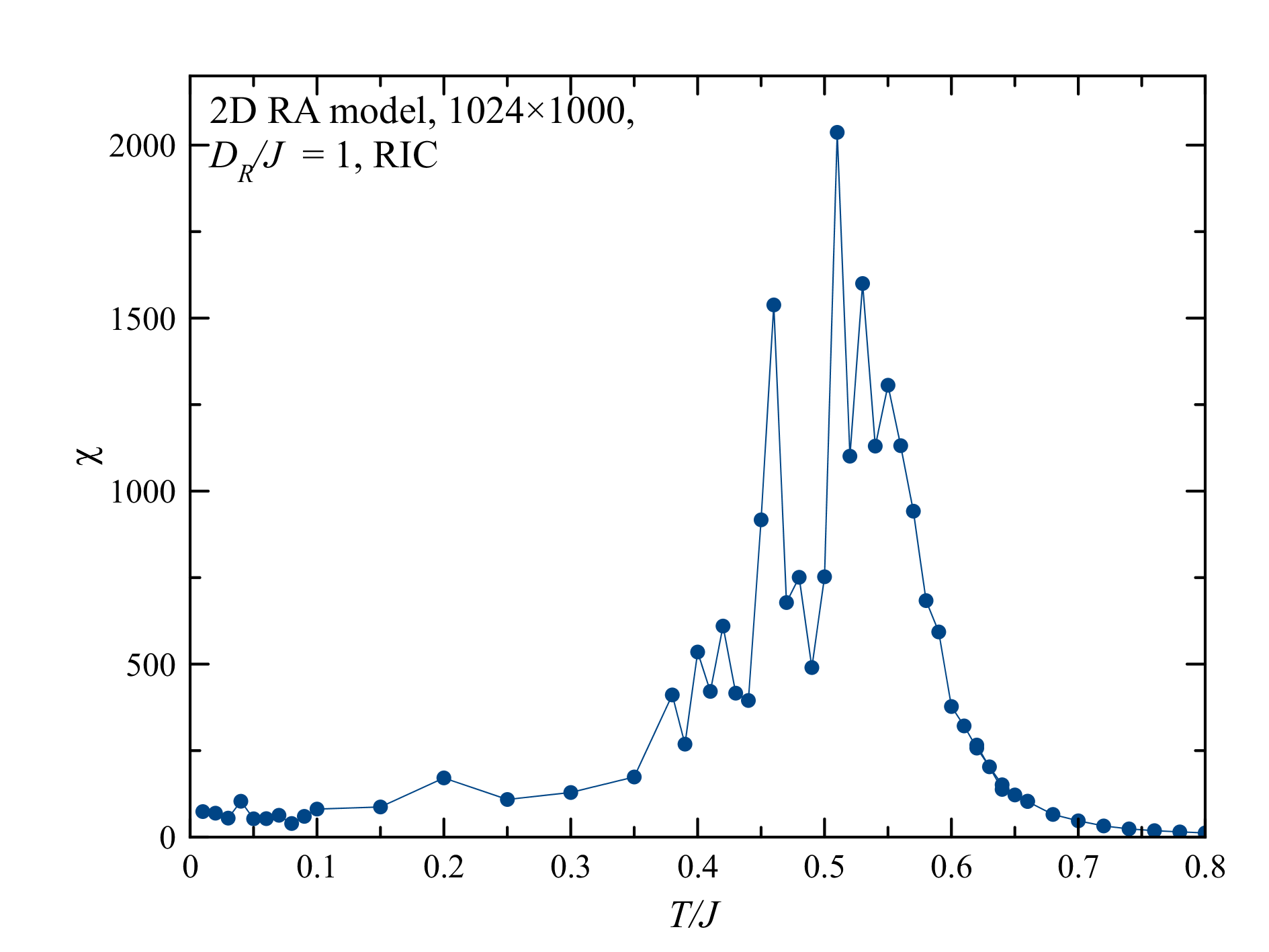}
\par\end{centering}
\begin{centering}
\includegraphics[width=9cm]{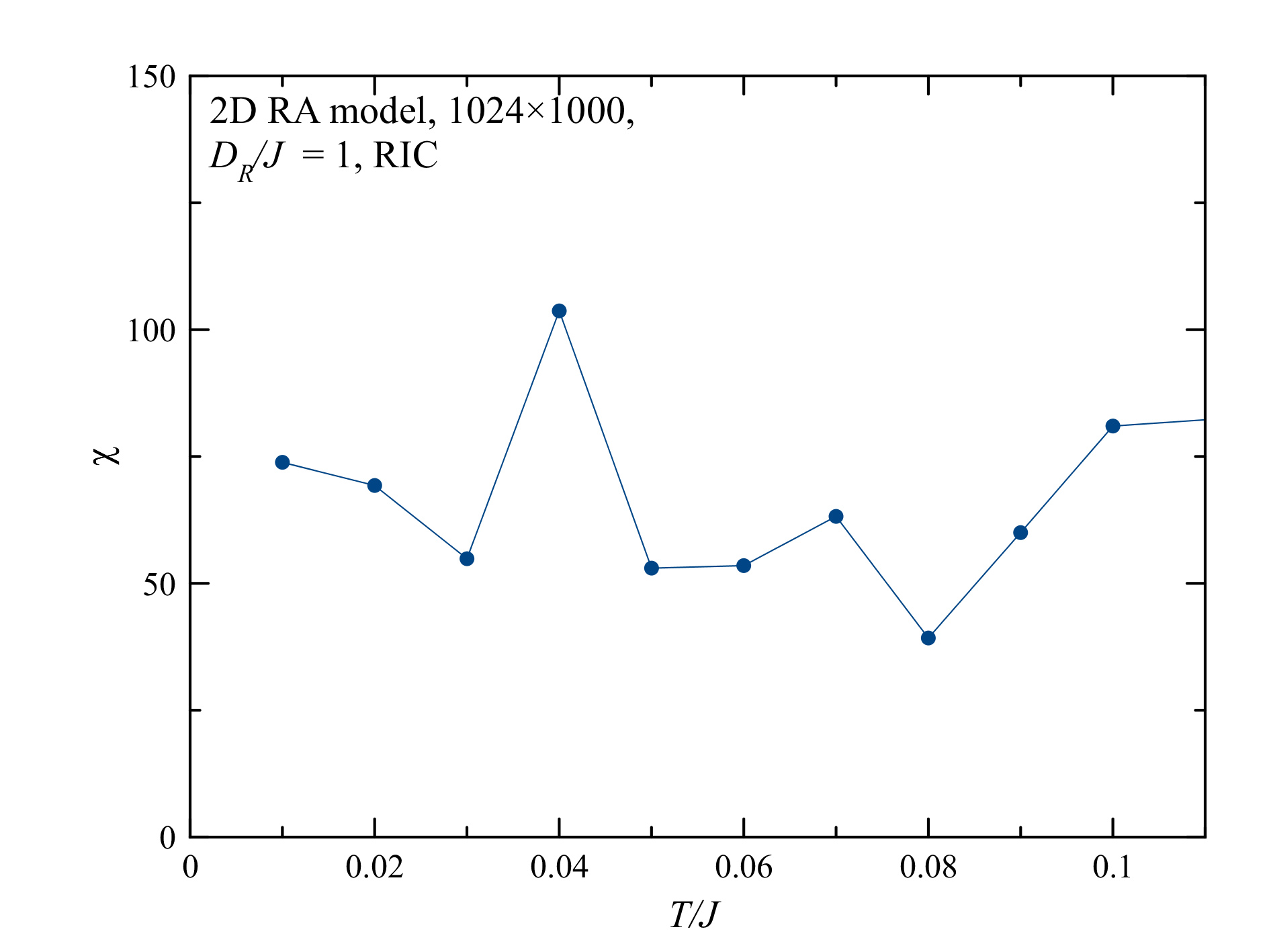}
\par\end{centering}
\caption{Linear susceptibility $\chi$ at different temperatures. Upper panel:
broad temperature range; Lower panel: Low temperatures.}

\label{Fig_chi_vs_T}
\end{figure}
Above the freezing point, the components of the magnetization defined
by Eq. (\ref{m_def}) fluctuate fast around zero, as can be seen in
the upper panel of Fig. \ref{Fig_T=00003D0.58}. Even for a disordered
system of one million spins the magnetization is significant that
is the consequence of a strong short-range order that establishes
below $T/J=0.7$ where the system has the maximum of the heat capacity,
even in the absence of the RA. Because of the fast fluctuations, the
linear susceptibility computed with the use of Eqs. (\ref{chi_comps})
and (\ref{chi_symm}) and shown in the lower panel of Fig. \ref{Fig_T=00003D0.58}
reaches its asymptotic value within the simulation interval of $10^{6}$
MCS. Different components of the susceptibility have approximately
the same value.

On the contrast, in the intermediate temperature range at and below
freezing, in addition to fast fluctuations of the magnetization, there
is slow dynamics, apparently due to thermally-activated barrier crossing.
As can be seen in the upper panel of Fig. \ref{Fig_T=00003D0.51},
slow changes of $\mathbf{m}$ do not average out within the simulation
interval of $10^{6}$ MCS. Slow fluctuations of $\mathbf{m}$ are
large and thus make a large contribution to the linear susceptibility.
As can be seen in the lower panel of Fig. \ref{Fig_T=00003D0.51},
the susceptibility does not stabilize and continues to grow. In this
temperature range, the susceptibility values are huge because of the
correlated motion of large groups of spins over energy barriers.

At lower temperatures, there are no large fluctuations of the magnetization
due to overbarrier transitions, as can be seen in the upper panel
of Fig. \ref{Fig_T=00003D0.3}. The fluctuations seen in the figure
are due to the motion of correlated spin bundles within their valleys.
Note that $z$ axis is chosen in the direction of the magnetization
in the initial state obtained by the energy minimization from the
random spin state.

The values of the symmetrized linear susceptibility $\chi$ at different
temperatures computed using Eq. (\ref{chi_symm}) are shown in Fig.
\ref{Fig_chi_vs_T}. At higher temperatures, the susceptibility is
small and practically coincides with that of the pure system. One
can see that $\chi$ has huge values in the region of freezing. However,
the values obtained in this region strongly fluctuate and are only
approximate as the simulation duration of $10^{6}$ MCS proves to
be insufficient to average out the magnetization fluctuations (see
the upper panel of Fig. \ref{Fig_T=00003D0.51}). Better results,
probably, could be obtained for the simulation longer by an order
of magnitude that for such a large system is problematic. At low temperatures,
the scatter in the susceptibility decreases and the susceptibility
values approach a plateau with the height in a fair accordance with
Eq. (\ref{chi_estimation}) that for $D_{R}/J=1$ with $R_{f}/a\approx13.3$
and $k=0.5$ yields $\chi J\simeq88$.

In addition, one could compute the correlation functions (CFs) of
the $\mathbf{m}$ time series shown in figures above. Above freezing,
these CFs quickly decay to zero, At intermediate temperatures, they
decrease slowly with large fluctuations, as suggested by the upper
panel of Fig. \ref{Fig_T=00003D0.51}. At low temperatures where there
are no overbarrier transitions, time CFs quickly decrease from their
equal-time values to their plateau values. These CFs have been computed
for the same model from the dynamical evolution and shown in Fig.
3 of Ref. \onlinecite{garchu_arXiv}, Those results suggest freezing
at $T/J$ between 0.5 and 0.6, in accordance with the current, more
precise, results.

\section{Discussion}

Most of the previous studies of random-anisotropy (RA) magnets were
focused on their equilibrium behavior or on their quasi-equilibrium
properties in a frozen glassy state. Rigorous analytical solution
of this set of problems has never been provided, while numerical studies
have been hampered by the necessity to consider large systems in order
to account for extended ferromagnetic correlations. Capabilities of
modern computers have allowed us to revisit this problem. In this
paper we have studied glassy properties of the random-anisotropy magnet
as a function of temperature with the combination of the Metropolis
Monte Carlo method and specially developed thermalized overrelaxation.

The questions we asked are the extent to which metastability plays
a role in defining magnetic properties of such a system, the freezing
of the magnetic configuration due to energy barriers on lowering temperature
vs a spin-glass transition due to exchange interaction between spins,
the time evolution of a conventionally defined spin-glass order parameter,
the characteristic melting time in the temperature region just above
freezing, the field-cooled (FC) and zero-field-cooled (ZFC) magnetization
curves, and the temperature dependence of the magnetic susceptibility
of RA magnets. The computed energy barriers agree within order of
magnitude with the Imry-Ma argument for systems with quenched disorder.
The computed FC and ZFC magnetization curves have close resemblance
with the experimental curves. These findings provide confidence in
our numerical method.

A more challenging task has been distinguishing between blocking of
overbarrier spin-group transitions on reducing temperature (that implies
the Arrhenius temperature dependence of the melting time above freezing)
and a true spin-glass phase transition (that implies a power-law divergence
of the melting time at transition point). While we cannot say with
confidence that we have answered this question unambiguously, our
findings provide a stronger argument in favor of a continuous freezing
(blocking) transition on lowering temperature. The main evidence of
this is a rather high power in the power-law fit of the melting time
and a rather low resulting transition temperature. In accordance with
our numerical experiments, the maximum of the susceptibility occurs
where the FC and ZFC magnetization curves merge. The low temperature
value of the susceptibility roughly agrees with the one derived from
the Imry-Ma argument. The temperature dependence of the susceptibility
near and below the maximum has a strong scatter caused by the finite
size of the system (one million spins) and finite computing time (one
million Monte Carlo steps). It did not allow us to distinguish between
a smooth behavior at the maximum and a cusp that was experimentally
observed in spin glasses. Studies of larger systems and longer computation
times would be needed to make such a distinction.

\section*{Acknowledgments}

This work has been supported by the Grant No. FA9550-20-1-0299 funded
by the Air Force Office of Scientific Research.

\end{document}